\documentclass[amsmath,onecolumn, floatfix,nofootinbib, prx, aps, usenames, dvipsnames,11pt]{quantumarticle}
\pdfoutput=1

\usepackage[utf8]{inputenc}
\usepackage[english]{babel}
\usepackage[T1]{fontenc}
\usepackage{natbib}

\usepackage[usenames,dvipsnames]{xcolor}
\usepackage{graphicx}
\usepackage[ colorlinks = true, 
             linkcolor = MidnightBlue,
             urlcolor  = MidnightBlue,
             citecolor = orange,
             anchorcolor = ForestGreen,
]{hyperref} 
\usepackage{float}

\usepackage[sans]{dsfont} 
\usepackage{bbm}
\usepackage[mathscr]{euscript}
\usepackage{cancel} 
\usepackage{amsfonts}

\usepackage{tcolorbox}
\tcbuselibrary{theorems} 

\usepackage{amsmath}
\usepackage{amssymb}
\usepackage{units}
\usepackage{framed}
\usepackage{mdframed}
\usepackage{thmtools, thm-restate} 
\usepackage{ucs}
\usepackage{subcaption}
\usepackage{fullpage}
\usepackage{csquotes}
\usepackage{epigraph}

\usepackage{multicol}
\usepackage{enumerate}

\usepackage{tikz}
\usetikzlibrary{arrows}
\usepackage{rotating, xcolor}
\usetikzlibrary{shapes}
\usetikzlibrary{hobby}
\usetikzlibrary{decorations.markings}

\makeatletter
\def\l@subsubsection#1#2{}
\def\l@subsection#1#2{}
\makeatother

\tcbset{colback=orange!5,colframe=orange!75!yellow, 
fonttitle= \bfseries, float=htb}

    \newcommand{\ket}[1]{\vert  #1 \rangle}
    
    \newcommand{\inprod}[2]{\langle #1 | #2 \rangle}

\newcommand{\footnoterecall}[1]{\hyperref[#1]{\footnotemark[\value{#1}]}}

\usepackage{cleveref}

\usepackage[page]{totalcount}
\usepackage{tcolorbox}

\usetikzlibrary{arrows,decorations.markings}
\usepackage{tikz-cd}
\usepackage{mathrsfs}
\usepackage{qcircuit}

\begin{document}

\title{Testing quantum theory with thought experiments}

\author{Nuriya Nurgalieva}
\affiliation{Institute for Theoretical Physics, ETH Zurich, 8093 Z\"{u}rich, Switzerland}
\email{nuriya@phys.ethz.ch}

\author{Renato Renner}
\affiliation{Institute for Theoretical Physics, ETH Zurich, 8093 Z\"{u}rich, Switzerland}
\email{renner@ethz.ch}

\begin{abstract}
Quantum mechanics is one of our most successful physical theories; its predictions agree with experimental observations to an extremely high accuracy. However, the bare formalism of quantum theory does not provide straightforward answers to seemingly simple questions: for example, how should one model systems that include agents who are themselves using quantum theory? These foundational questions may be investigated with a theorist's tool – the thought experiment. Its purpose is to turn debates about the interpretation of quantum theory into actual physics questions. In this article, we give a state-of-the-art overview on quantum thought experiments involving observers, from the basic Wigner's friend to the recent Frauchiger-Renner setup, clarifying their interpretational significance and responding to objections and criticism on the way.
\end{abstract}

\maketitle


\setlength{\epigraphwidth}{4in}
\epigraph{Imagination will often carry us to worlds that never were. But without it we go nowhere.}{Carl Sagan, \emph{Cosmos (1980)}}

\section{Introduction}
\label{sec:introduction}


\newcommand{\RR}[1]{{\bf [RR: #1]}}

\newcommand{\states}[2]{\ket{\text{{\footnotesize ``#1''}}}_{#2}}
\newcommand{\statesn}[2]{\ket{\text{{\small #1}}}_{#2}}

The development of quantum physics started in the early twentieth century as a response to a number of experimental observations which did not find an explanation in the realm of classical physics. The result was a new fundamental and rather universal theory of nature. The formalism of quantum theory allows us to describe essentially any phenomena that does not involve gravity, from the photoelectric effect to superconductivity in condensed matter physics and to the generation of new particles by collisions in high-energy physics. The predictions of the theory agree with experimental observations to an extremely high accuracy. Because tests of quantum properties require almost perfectly isolated systems, these experiments were initially limited to rather simple setups on small scales. However, with the development of quantum technologies, more and more complex systems are investigated, and so far no indications have been found that quantum theory could be inaccurate on larger scales. 

Despite this success, physicists still disagree on seemingly elementary questions. Does the observation of the output of a quantum random number generator lead to a collapse of its wave function~\cite{vonNeumann1955}?  And, if yes, when and why does this collapse happen?\footnote{These questions are often referred to as the \emph{measurement problem}~\cite{StanfordQuantum}.}  The bare formalism of quantum theory does not provide conclusive answers. It thus needs to be \emph{interpreted}. Many proposals for such interpretations were made in the course of a century~\cite{Bohr1934,Heisenberg1958,Bohm1952,Everett1957,Brassard2019, Wallace2002,Rovelli1996, Caves2002, Froehlich2015,Griffiths1984, Gellmann1997,Ghirardi1985, Gambini2015,BaumannWolf2018}; however, no consensus has been reached yet. 

Interpretations are, roughly speaking, attempts to give physical meaning to the  mathematical objects quantum theory talks about. What does it mean, for instance, if a calculation within the formalism of quantum theory shows that the observer of the output of a quantum random number generator becomes entangled with the generator? Questions like these do not arise in classical theories, for the quantities appearing in the formalism ---  locations and velocities of objects --- naturally correspond to ontological and observable facts.   


For many years, the debate on the interpretational meaning of quantum theory was perceived as purely philosophical and not providing any significant physical insights. However, this is changing in the light of recent theoretical research, which investigates the consistency of interpretations via a novel type of thought experiments.  Their purpose is to turn questions about the interpretation of quantum theory into actual physics questions. Describing these thought experiments and their physical implications is the main aim of this review. 

Thought experiments have a long tradition in physics as sources of inspiration and conceptual insight.   A prominent early example is the \emph{Maxwell's demon paradox}~\cite{Maxwell1871}. Here one considers a little agent, the ``demon'', who is able to observe and control the individual particles of a gas. Making use of this control, the demon may decrease the entropy of the gas --- in apparent violation of the second law of thermodynamics. The investigation of this paradox ultimately led to the insight that any irreversible processing of information (e.g., by the demon) has an unavoidable energy cost~\cite{Landauer1961, Bennett1987}. This is nowadays known as \emph{Landauer's principle}, and forms a bridge between information theory and thermodynamics (see~\cite{ThermoInfoRelation} and references therein). 

A remarkable aspect of the Maxwell's demon paradox is that the demon admits two different roles. It is on the one hand an agent who measures and acts upon the gas particles. On the other hand, to resolve the paradox it is crucial to regard the demon itself as a physical system that is subject to the laws of thermodynamics. This aspect occurs again in the famous \emph{Wigner's friend experiment}, proposed by Eugene Wigner in 1961 to shed light on the quantum measurement process~\cite{Wigner1961}. 

Wigner imagined  an agent, Alice (his ``friend'') who is located in a lab and measures a quantum system $R$. He assumed that $R$ is prepared in a  state of the form
\begin{align} \label{eq_SystemInitial}
   \ket{+}_R =  \sqrt{\frac{1}{2}} \bigl( \ket{0}_R+\ket{1}_R \bigr)
\end{align}
and that Alice measures it in the basis $\{\ket{0}_R,\ket{1}_R\}$. (Following  quantum information terminology, we will sometimes call this the \emph{computational basis}.) According to quantum theory, Alice will observe a definitive outcome~$a$, either $a=0$ or $a=1$, each with probability $1/2$. 

The same experiment may now be described from the viewpoint of an outside agent (Wigner), who chooses to regard Alice ($A$) and the measured system ($R$), as well as the entire environment ($E$), as a big quantum system.  Because this big system does not interact with anything else (we just defined it to include the entire environment), its time evolution is, according to quantum theory, reversible. Hence assuming that the experiment starts with $A$ and $E$ initialised to a pure state, the joint state of $A$, $R$, and $E$ after the measurement will still be pure and be of the form
\begin{align} \label{eq_WignerFriendState}
 \sqrt{\frac{1}{2}} \Bigl(\ket{0}_R\statesn{Alice concluded $a=0$}{A}\ket{\text{env}_0}_{E} + \ket{1}_R \statesn{Alice concluded $a=1$}{A}\ket{\text{env}_1}_{E} \Bigr) \ .
\end{align}

Note that, like Maxwell's demon, Alice plays two different roles here. On the one hand she is an agent who observes the measurement outcome $a$. On the other hand she is part of a big quantum system that evolves unitarily and may admit a state like~\eqref{eq_WignerFriendState}, in which both possible values for $a$ appear symmetrically. Both descriptions are compatible with the formalism of quantum theory --- the formalism itself does not provide criteria to decide which of the two is valid, if not both, nor what expressions like~\eqref{eq_WignerFriendState}  mean physically. The thought experiment of Wigner's friend, maybe along with Schr\"odinger's cat~\cite{Schroedinger1926}, is thus a prime example pointing at the need for an interpretation of the quantum formalism.

\begin{figure}[h]
\centering
\includegraphics[scale=0.3]{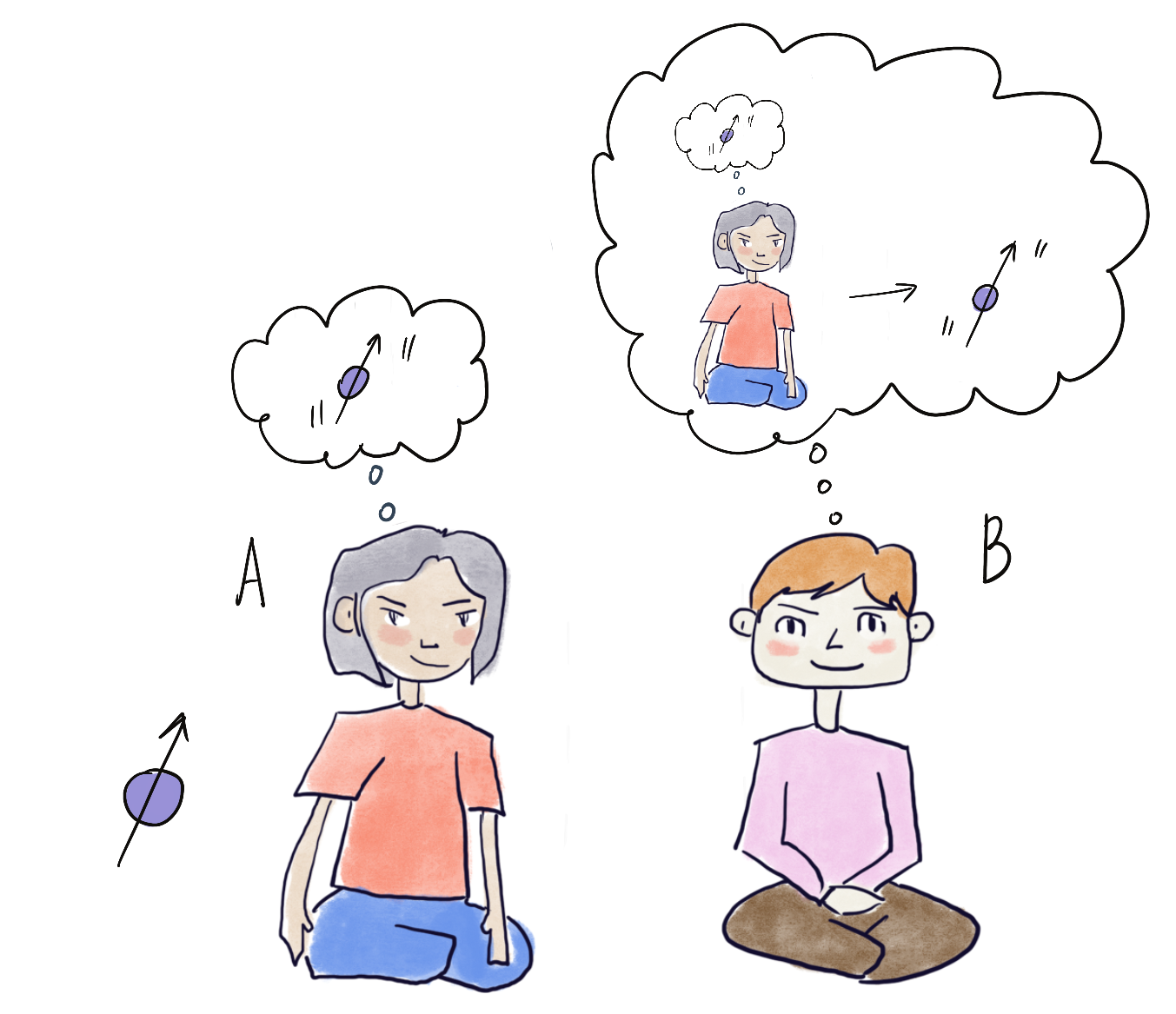}
\caption{{\bf Multi-agent experiments as consistency tests for physical theories.} If a proposed physical theory is supposed to be universal then it must  be possible to model agents, i.e., users of the theory, as physical systems within that theory. In other words, the theory must be able describe its own users! This immediately raises questions of consistency. For example, an agent Bob may not only reason directly about a particle, but he may as well use the theory to describe how another agent Alice reasons about the same. Within a consistent theory, such indirect reasoning should be allowed and not lead to contradictions.}
\label{fig:teaser}
\end{figure}

Wigner's thought experiment was followed by a number of extensions --- by David Deutsch in 1985~\cite{Deutsch1985}, and more recently by \v Caslav Brukner~\cite{Brukner2018}, and by Daniela Frauchiger and one of us~\cite{Frauchiger2018} (in the following referred to as the \emph{FR thought experiment}).\footnote{Preliminary versions of these more recent works appeared in \href{https://arxiv.org/abs/1507.05255v1}{arXiv:1507.05255v1} and \href{https://arxiv.org/abs/1604.07422v1}{arXiv:1604.07422v1}, respectively.} These serve as a metaphorical lantern helping to navigate the landscape of different interpretations of quantum theory, identify key elements which make them distinct, and explore their consistency (see Figure~\ref{fig:teaser}). But, even more importantly, these thought experiments show that choosing an interpretation is not a matter of philosophy but a matter of physics. As we shall see, when asked to predict the experiments' outcomes, the answers generally depend on that choice!

This article is structured as follows. As is apparent from the introduction above, the notion of an \emph{observer} (or \emph{agent}) plays a key role, and we thus start with a discussion of what various interpretations have to say about it (Section~\ref{sec:observer}). We then proceed with a description of increasingly sophisticated thought experiments, starting with Wigner's, which we already mentioned briefly (Section~\ref{sec:wigner}), its modification by Deutsch (Section~\ref{sec:deutsch}), and finally the FR thought experiment (Section~\ref{sec:fr}).\footnote{This thought experiment has many similarities to the thought experiment proposed by Brukner~\cite{Brukner2018}. The conclusions drawn from the two thought experiments are however rather different. Brukner's experiment has led to an important strengthening of Bell's theorem (see Footnote~\ref{ftn:ObserverIndependentFacts} and~\cite{Cavalcanti}). Conversely, the FR thought experiment may be regarded as a test of whether quantum theory is able to consistently describe its own users, as illustrated by Figure~\ref{fig:teaser}. This is the main theme here and we therefore focus on the latter.} We then address some of the most frequent questions that arise in debates about these thought experiments (Section~\ref{sec:criticism}). We conclude with a discussion of broader implications and an outlook on future research (Section~\ref{sec:discussion}).

\section{Observers in quantum mechanics}
\label{sec:observer}
The notion of an \emph{observation} is crucial for linking the theoretical formalism of quantum theory to experiment, and in this sense to physical reality. In an experiment, an observation is the outcome of a measurement that is carried out by an \emph{observer}, which may be a device or a human. How this observer should be treated in theory is however debated. Is the observer just an ordinary system that can itself be described by quantum theory, or should it be regarded as something external to it, or is it even both? The answer to this question exhibits some of the key differences between the different interpretations of quantum theory, and we thus discuss it in more detail in this section. 

The notion of an observer gained importance in physics already before the development of quantum theory. One of its most prominent appearances is in Einstein's original article on special relativity~\cite{Einstein1905}. Here the ``Beobachter'' plays a key role, for quantities such as time and spatial locations are only defined relative to them. Special and general relativity still portray the observer in a passive light --- the observer is understood as a reference frame, thus giving rise to a coordinate system in spacetime \cite{Doran2003, Nerlich1994}. Other theories view observers as users of the theory. For example, Bayesian statistics~\cite{Cox1946, Jaynes1970} consists of a set of reasoning rules that a \emph{rational agent} should follow to make predictions and decide on future actions based on past observations. A Bayesian observer thus takes a more active role. In quantum theory, both of these roles may be relevant, depending on the interpretation. We will therefore use the terms \emph{agent} and \emph{observer} interchangeably.

\begin{figure}[p]
\centering
\includegraphics[scale=0.25]{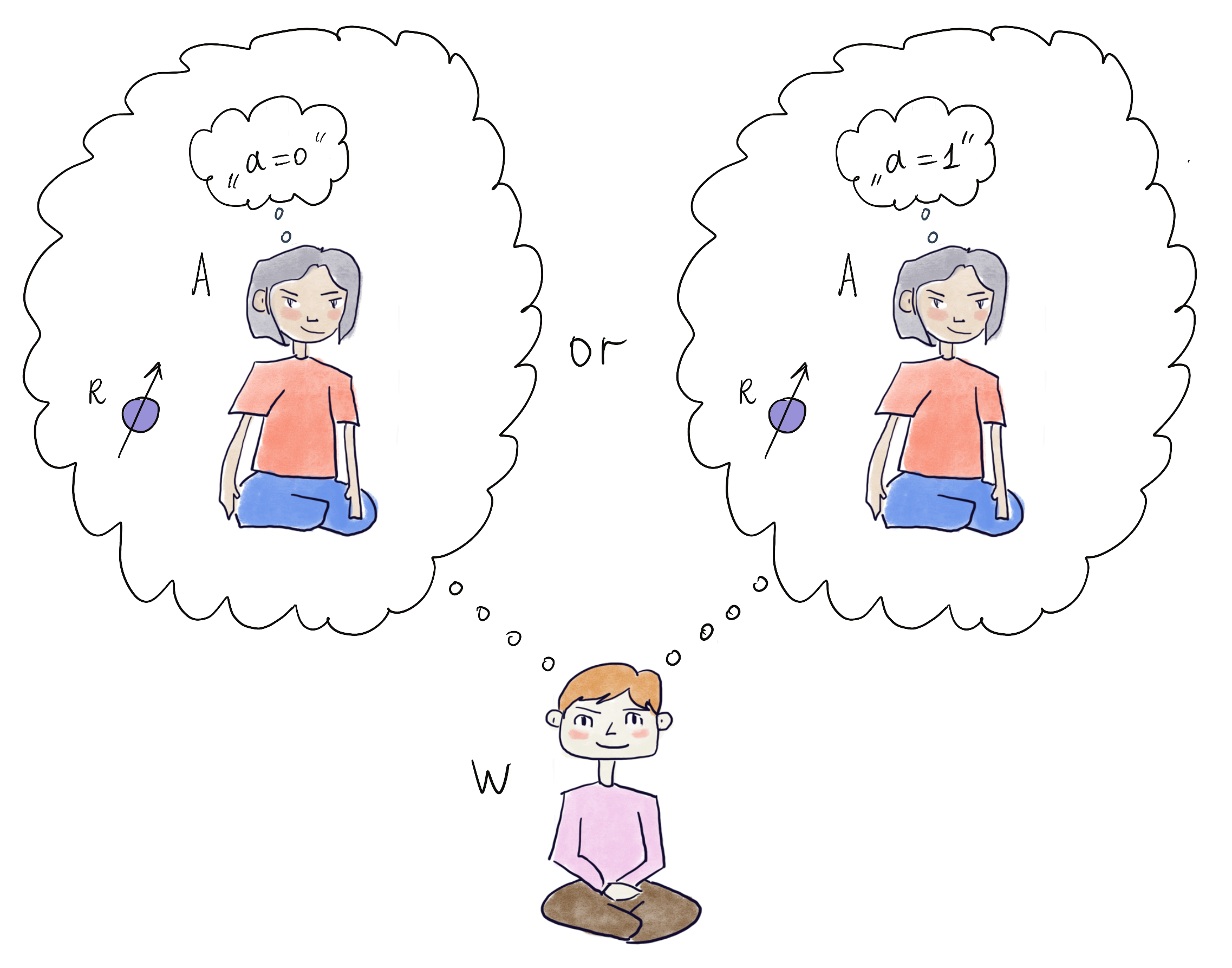}
\caption{{\bf Alice and Wigner's views.} After Alice has completed her measurement, the external agent Wigner may conclude that Alice, together with the entire environment, is in a superposition state like~\eqref{eq_WignerFriendState}, consisting of two equal branches, one in which she has observed $a=0$ and one in which she has observed $a=1$. But doesn't Wigner's conclusion contradict Alice's view, whom we would expect to have observed only one of these values? The answer largely depends on the roles one assigns to the different observers. For example, an interpretation may assert that, if Alice is part of a system that is in a superposition state, then she cannot be considered a legitimate observer, and there is hence no point in talking about Alice's view.}
\label{fig:wignerIntro}
\end{figure}

The Wigner's friend thought experiment is an excellent test ground to explore the different roles that are given to observers by the different interpretations of quantum theory (Figure~\ref{fig:wignerIntro}). Wigner's idea was to study a scenario where ``one does not make the observation oneself but lets someone else carry it out''~\cite{Wigner1961}. As explained in the introduction, he considered an agent (Alice) who is measuring a system prepared in a superposition state, and he concluded that from an outside agent's (Wigner's) perspective, the measurement process is according to the formalism of quantum theory described by a reversible state transformation that may result in a superposition state like~\eqref{eq_WignerFriendState}. 

Wigner regarded this as a contradiction. He argued that, clearly, Alice observes one outcome, $a=0$ or $a=1$, and thus the joint state of the measured system ($R$), Alice ($A$), and the environment ($E$) after the measurement should be either
\begin{align*}
    \ket{0}_R\statesn{Alice concluded $a=0$}{A}\ket{\text{env}_0}_{E}
\end{align*}
or
\begin{align*}
    \ket{1}_R\statesn{Alice concluded $a=1$}{A}\ket{\text{env}_1}_{E} \ ,
\end{align*}
but not a linear combination of the two, which ``appears absurd because it implies that my friend was in a state of suspended animation''~\cite{Wigner1961}. Wigner concluded that a conscious observer is not correctly described by the formalism of quantum theory.

\begin{figure}[p]
\centering
    \begin{subfigure}{0.45\textwidth}
        \includegraphics[scale=0.17]{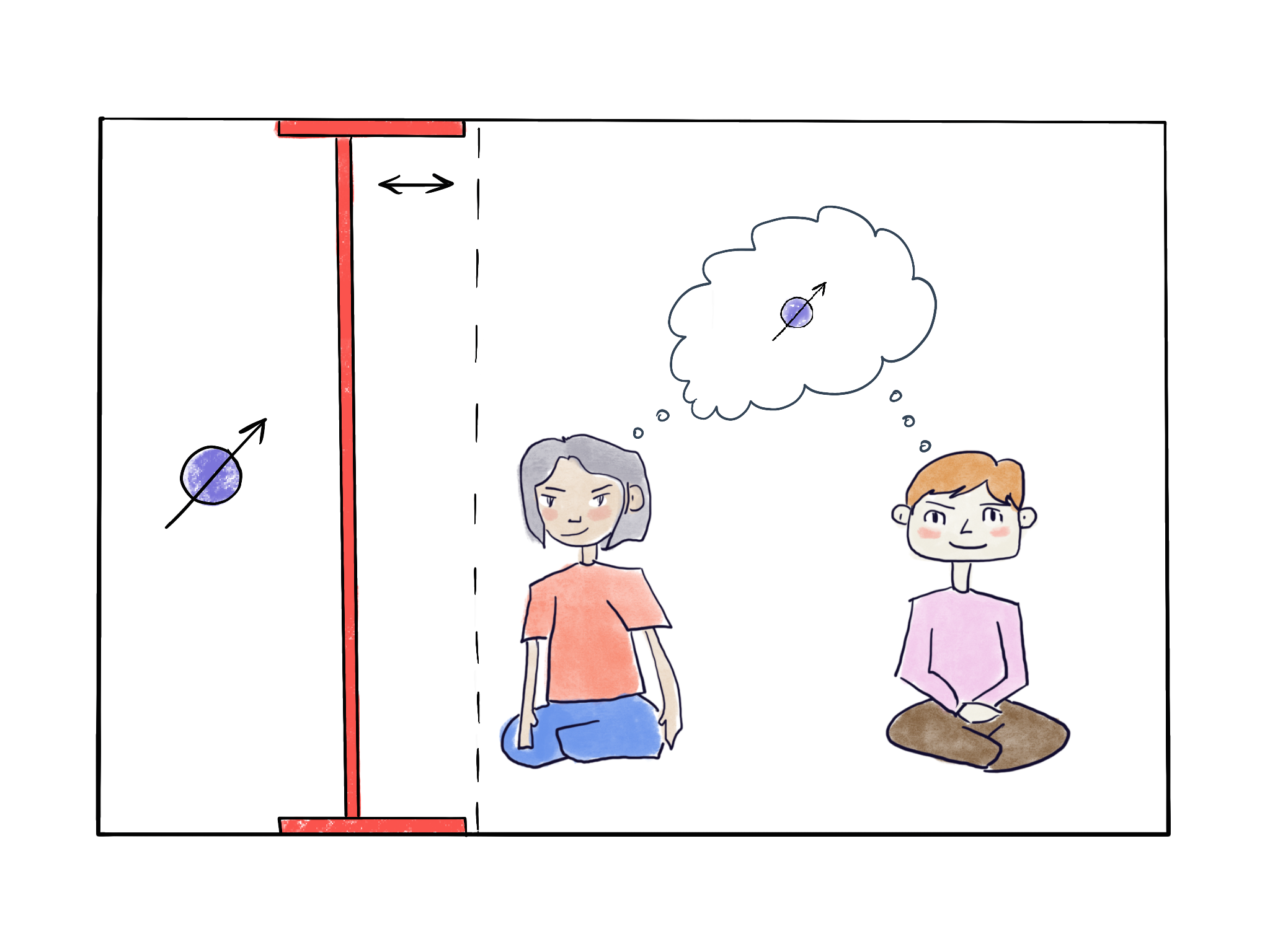}
        \caption{\textbf{Conventional Copenhagen.} The cut is the same for all observers, and they all need to be placed above it.}
        \label{fig:copenhagen}
    \end{subfigure}
    \
    \begin{subfigure}{0.45\textwidth}
        \includegraphics[scale=0.17]{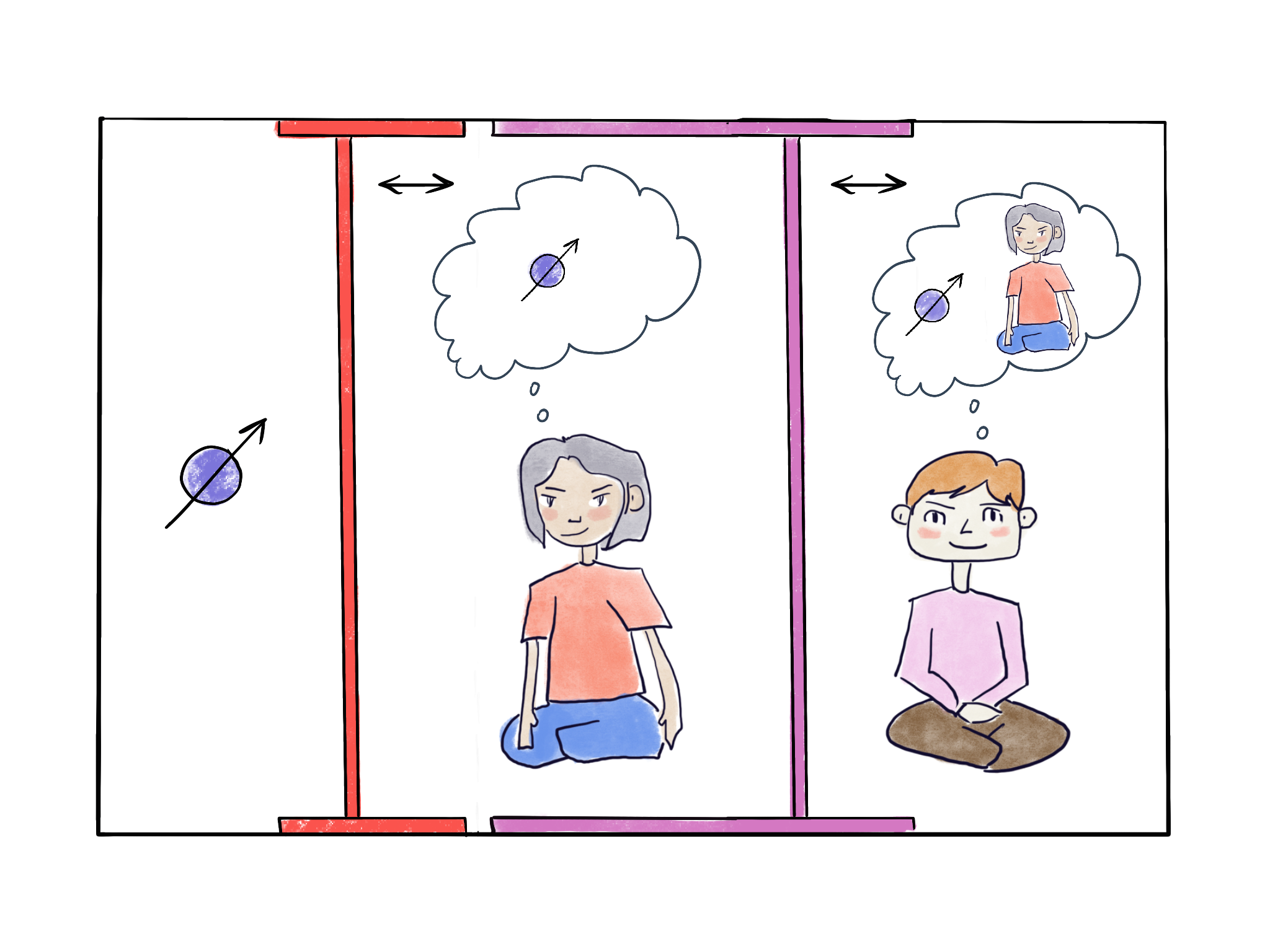}
        \caption{\textbf{Neo-Copenhagen.} The cut is subjective and motile: every observer chooses it relative to themselves but must remain above it.}
        \label{fig:neo}
    \end{subfigure}
    \\
    \begin{subfigure}{0.45\textwidth}
        \includegraphics[scale=0.17]{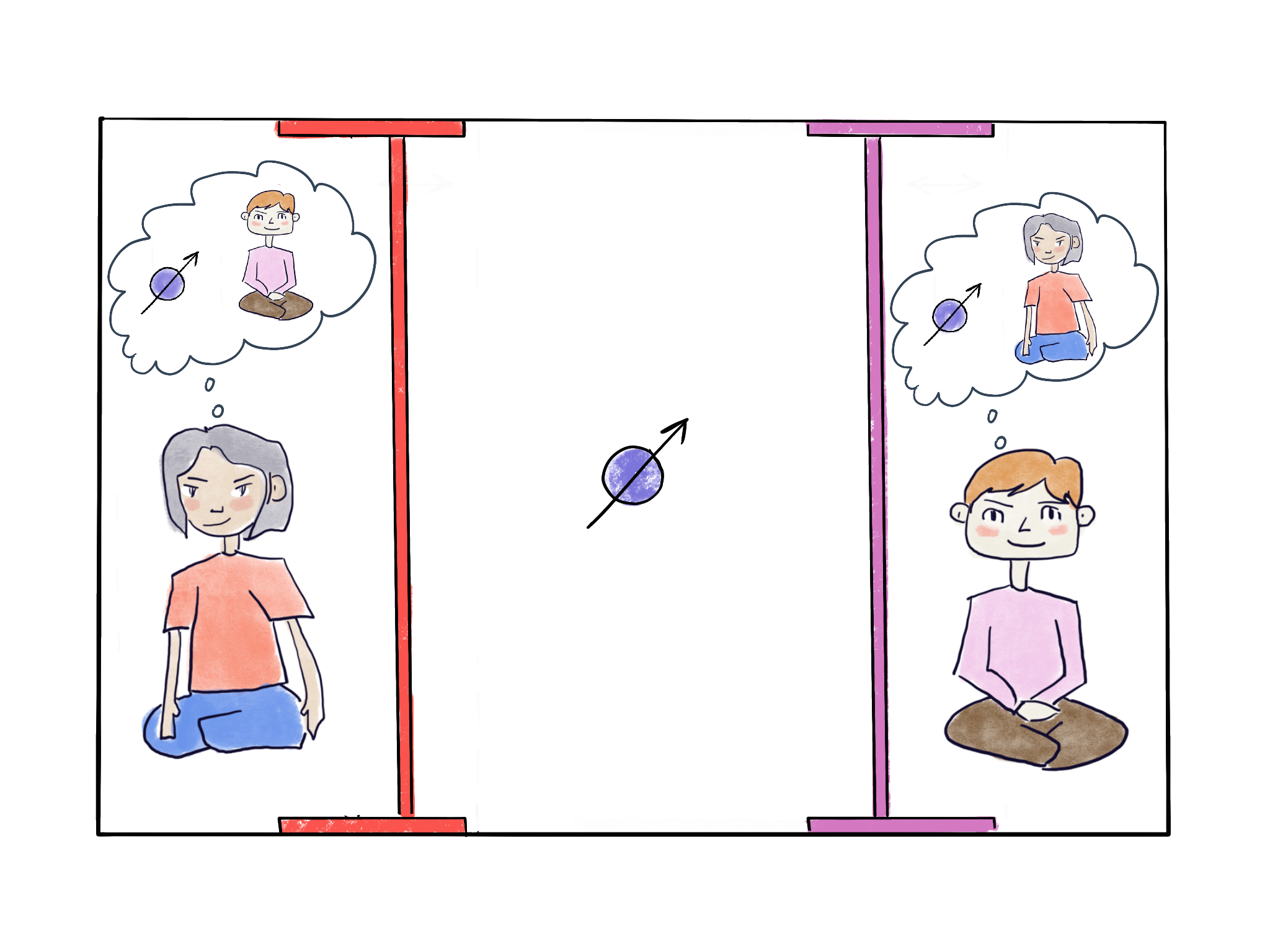}
        \caption{\textbf{QBism.} Any observer puts themselves above their personal cut and everything else below it.}
        \label{fig:qbism}
    \end{subfigure}
    \
    \begin{subfigure}{0.45\textwidth}
        \includegraphics[scale=0.17]{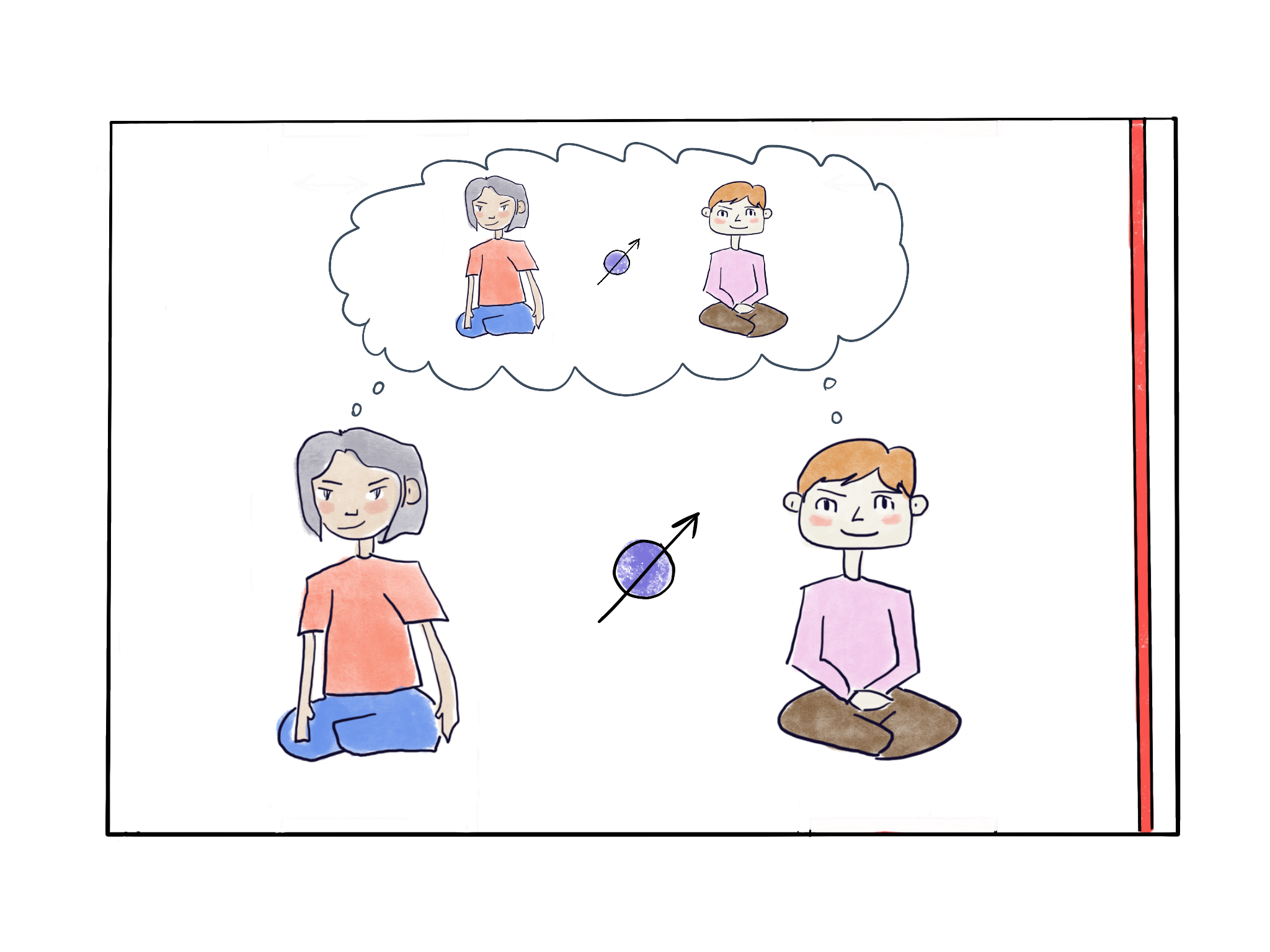}
        \caption{\textbf{Many-worlds and Bohmian mechanics.} The entire universe, including any observer, is below the cut.}
        \label{fig:manyworlds}
    \end{subfigure}
    \\
    \begin{subfigure}{0.45\textwidth}
        \includegraphics[scale=0.17]{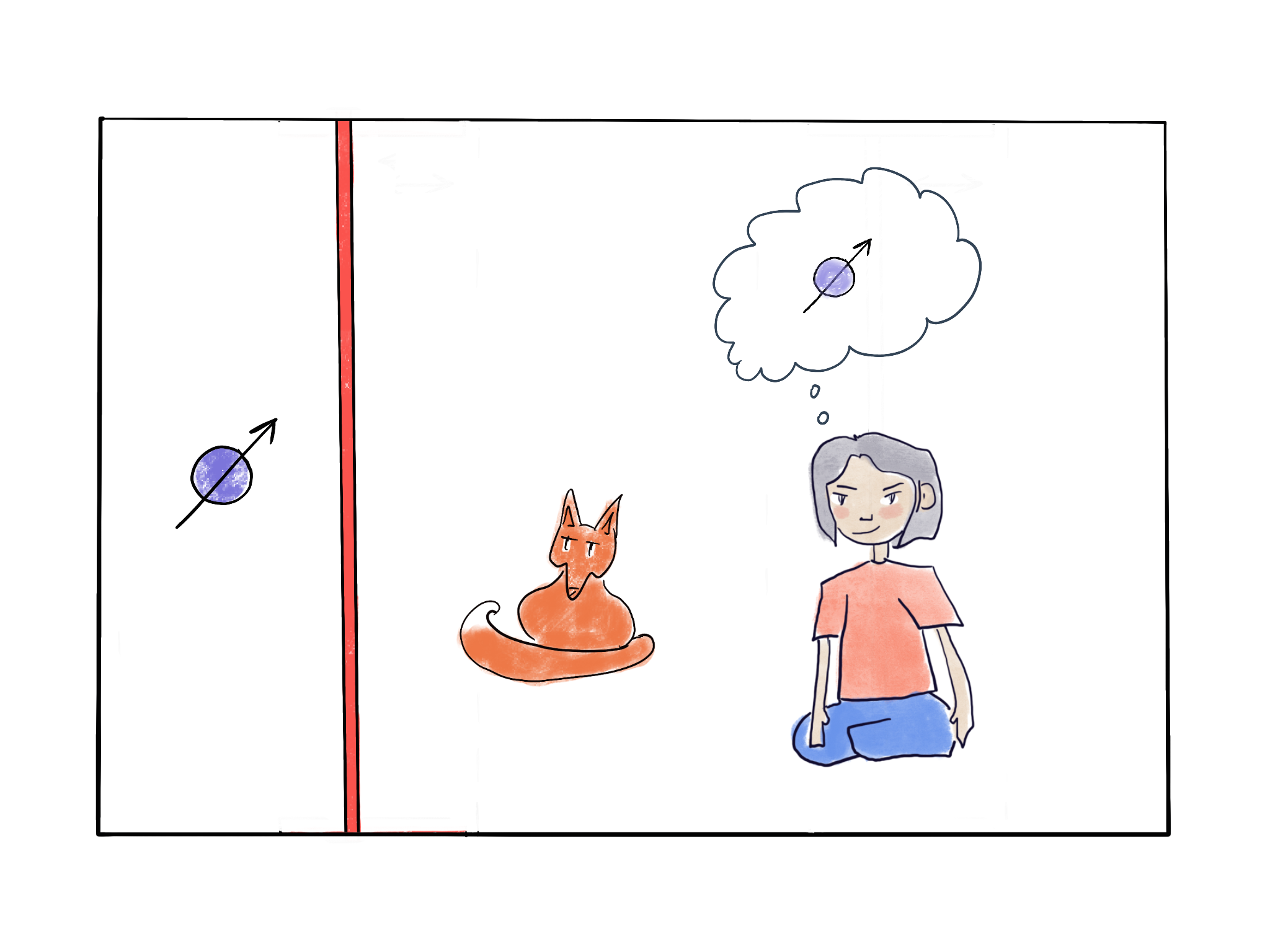}
        \caption{\textbf{Collapse theories.} The cut is implied by an extra decoherence mechanism, which affects macroscopic systems but not microscopic ones.}
        \label{fig:collapse}
    \end{subfigure}
\caption{\textbf{The Heisenberg cut in different interpretations of quantum theory.} The cut indicates a conceptual separation between subsystems of the universe. Systems above or below the cut are regarded as classical or quantum, respectively. The cut may be objective or subjective. In the latter case its colour matches the agent relative to which it is defined.}
\label{fig:cut}
\end{figure}

While most physicists today don't resort to the (vague) notion of consciousness,\footnote{Also Wigner changed his mind later and took the stance that  quantum theory is generally not valid for macroscopic systems, independently of whether they have consciousness or not (see~\cite{Esfeld1999}).} they still haven't reached an agreement on how to resolve the paradox around Wigner's friend. The ambiguity is not \emph{within} the formalism of quantum theory, but lies in \emph{how} the formalism should be applied to such experiments --- a question that the formalism itself cannot answer.
In their search for answers, quantum physicists came up with a variety of ideas and proposals, which are nowadays known as  different interpretations of quantum theory  (see Table~\ref{tab:observer} for a summary and Figure~\ref{fig:cut} for an illustration).

\subsection{Copenhagen interpretations}
\label{subsec:copenhagen}

The \emph{Copenhagen interpretation} \cite{Bohr1934,Bohr1958,Heisenberg1958,StanfordCopenhagen} answers the question about the role of the observer as follows. For the practical purpose of reasoning about an experiment, we must conceptually split the experiment into two parts. One part is the observed system, which is described in the language of quantum mechanics. The other part is the observing system, which consists of the measuring device and the agent that acquires knowledge about the measurement outcome. This second part must be described in ordinary (classical) language. The  observer thus should be able to say, for example, ``I started the measurement process and I observed the outcome $a=0$.''

This split of the world into two different parts is called the \emph{Heisenberg cut}~\cite{Atmanspacher1997}. Above the cut one uses classical descriptions, while the part below the cut forms the actual quantum system. The Heisenberg cut is movable to some extent. For example, one may shift it from below the measuring device to above it, thus including the device in the quantum domain. However, the measuring agent must always remain within the classical domain. The Copenhagen interpretation hence disallows the existence of a ``wave function of the universe''. 

The Copenhagen interpretation itself still leaves room for interpretation, and there are hence several sub-interpretations. In particular, the Heisenberg cut can be regarded \emph{objective} or \emph{subjective}. In the subjective variant, any agent that uses quantum theory may choose individually where to place her personal cut, under the sole constraint that she remains herself in the classical realm. This view has become increasingly popular in recent years and is often referred to as the \emph{neo-Copenhagen} interpretation~\cite{Brukner2017,deMuynck2007}. Conversely, we will use the term \emph{conventional Copenhagen} for the objective variant. Here the Heisenberg cut is the same for all agents.

\begin{tcolorbox}
\textbf{Observers in conventional Copenhagen} \\
The Heisenberg cut, which separates reality into a classical and a quantum part, is understood as objective, i.e., the same for all agents. According to this original understanding of the Copenhagen interpretation,  \emph{all} observers must be placed on the classical side of the cut.
\end{tcolorbox}
\label{subsubsec:objCop}

\begin{tcolorbox}
\textbf{Observers in neo-Copenhagen} \\
The Heisenberg cut is viewed as subjective, i.e., it depends on the agent who is applying the quantum formalism. Any agent can thus individually decide where to place the cut, provided that the agent themselves remains in the classical domain. In multi-agent scenarios, however, an agent may place other agents into the quantum domain.
\end{tcolorbox}

The conventional Copenhagen interpretation does not provide clear instructions as to where to place the cut in the Wigner's friend thought experiment. If one decides to put it above Alice then she is part of the quantum domain, which may legitimately be in a superposition of states corresponding to $a=0$ and $a=1$. However, in this case Alice can no longer be regarded as an agent who has observed a definitive measurement outcome, i.e., claiming that Alice observed either $a=0$ or $a=1$ is disallowed. Conversely, if one decides to place the cut below Alice then she is an observer, and her measurement yields a definitive outcome, $a=0$ or $a=1$. But since Alice is now in the classical domain, talking about a superposition state like~\eqref{eq_WignerFriendState} that involves Alice as a subsystem has no meaning.

This is different in the neo-Copenhagen interpretation, where the Heisenberg cut is subjective and hence agent-dependent. Here Alice would naturally place the cut somewhere in between the measured system and herself. It is thus legitimate to say that, from Alice's viewpoint, the measurement has a definitive outcome, $a=0$ or $a=1$. At the same time, Wigner may regard Alice as a quantum system. Expression~\eqref{eq_WignerFriendState} thus makes sense according to neo-Copenhagen as a subjective description from Wigner's viewpoint. In summary, neo-Copenhagen yields two different descriptions of the same experiment, each corresponding to one of the agents' viewpoints, and it is accepted that these viewpoints can be different. 

\subsection{Bohmian mechanics}
\label{subsec:bohmian}

The \emph{pilot-wave approach} of de Broglie and Bohm, also known as \emph{Bohmian mechanics}~\cite{Bohm1952, StanfordBohmian}, completely avoids the  Copenhagen interpretation's separation of the experiment into a classical and a quantum realm, i.e., there is no Heisenberg cut. Rather, according to Bohmian mechanics,  the description of an experiment --- in fact the entire universe --- consists of a global quantum state, the ``wave function of the universe'',  \emph{and} a classical configuration, whereby the latter determines our actual observations.  

Specifically, the classical configuration of the universe is described by the coordinates of all particles. Their dynamics is governed by a \emph{guiding equation}, which determines each particle's velocity. This equation also depends on the wave function of the universe, which in turn evolves according to the Schr\"odinger equation. One may thus think of each particle as a surfer: it occupies a particular place at any moment in time, yet its motion is dictated by a spread-out wave\footnote{This is is not just a wave in the $3$-dimensional real space, but in  $3^N$-dimensional Hilbert space, where $N$ is the number of particles one considers.}. According to the \emph{equilibrium hypothesis}~\cite{Durr1992}, the initial configuration of the particles is distributed  such that, under suitable typicality conditions, the probability of finding them somewhere is given by the Born rule applied to the wave function of the universe. (That is, the probability corresponds to the modulus squared of the wave function.) The guiding equation conserves this property as the system evolves in time. In this sense, Bohmian mechanics is, at least in typical situations, compatible with the formalism of quantum theory, applied to the universe as a whole. 

Bohmian mechanics may be regarded as a concrete example of a \emph{deterministic hidden variable theory} in the sense of Bell~\cite{Bell1966}, with the hidden variables corresponding to the classical configuration.\footnote{This terminology may however be misleading, for the classical configuration corresponds to actual observations of particle positions, whereas  the global quantum state remains hidden.} According to Bell's theorem, any such theory must be non-local. This is indeed the case for Bohmian mechanics, in the following sense: In an experiment involving multiple particles, the classical position and velocity of one particle may be influenced instantaneously by a manipulation of another remote particle.

\begin{tcolorbox}
\textbf{Observers in Bohmian mechanics} \\
Bohmian mechanics is a ``theory of the universe'' rather than a theory about particular systems. In particular, no distinction is made between observing systems and observed systems, and hence the notion of an observer plays no fundamental role. A user of the theory is thus urged to always take an outside perspective and reason about the wave function of the entire universe (which includes themselves). 
\end{tcolorbox}

A description of the Wigner's friend experiment within Bohmian mechanics would require the modelling of everything, including the friend, as a big quantum system. Since the time evolution of its wave function is governed by the Schr\"odinger equation, it will evolve towards a superposition state like~\eqref{eq_WignerFriendState}, with one branch corresponding to the friend observing the outcome $a=0$ and one corresponding to $a=1$. Still, according to the theory, the experiment always has a well-defined classical configuration, corresponding to one of the two outcomes. Furthermore, the evolution is deterministic. That is, given the configuration of all particles  at the beginning of the experiment, Bohmian mechanics would allow one to predict whether the friend will observe $a=0$ or $a=1$.

\subsection{Many-worlds interpretations}
\label{subsec:everett}

Everett's \emph{many-worlds interpretation} assumes that the state of the universe is completely described by a global wave function that evolves unitarily~\cite{Everett1957, StanfordMW}. The conscious experience that we observe a definite outcome upon measuring a system is considered a subjective illusion --- there is nothing in the theory that would reflect that outcome.  Rather, a measurement is regarded as a particular unitary evolution that leads to a \emph{branching} of the state of the universe. This means that the resulting wave function can be written as a superposition of terms, the \emph{branches}, each corresponding to one of the possible measurement outcomes. In subsequent work, the different branches have sometimes been termed \emph{parallel worlds}. 

\begin{tcolorbox}
\textbf{Observers in many-worlds} \\
Many-worlds interpretations describe the entire universe as a big quantum system, which necessarily includes any observer. Observers nonetheless have a role: their possible observations are the basis for defining the branching structure of the global quantum state.   
\end{tcolorbox}

Everett's original many-worlds interpretation is non-local, in the sense that  the branching imposed by a measurement affects the whole universe instantaneously. But there are also locally realistic variants of the interpretation, where the branching is restricted to locations to which the outcome of a measurement has been communicated. This approach, which was inspired by~\cite{Deutsch2000}, is known as \emph{parallel lives}~\cite{Brassard2019}. Additionally, a relativistic extension of the many-worlds interpretation was proposed in~\cite{Wallace2002}.

Applied to the Wigner's friend experiment, the many-worlds interpretation asserts that, after Alice's measurement, the universe is in a superposition state of the form~\eqref{eq_WignerFriendState}, consisting of one branch in which Alice has observed $a=0$ and one in which she has observed $a=1$. The global wave function is thus the same as in a Bohmian mechanics description of the experiment. However, in contrast to the latter, many-worlds explicitly does not break the symmetry between the different possible observations. While Alice's conscious perception is that of living in one single branch, and thus observing either $a=0$ \emph{or} $a=1$, both branches have the same ontic status.

\subsection{Relational quantum mechanics}
\label{subsec:relational}

In \emph{relational quantum mechanics (RQM)}~\cite{Rovelli1996,StanfordRQM} the state assigned to a system must always be understood as relational, in a similar sense as in relativity theory, for instance, the $3$-velocity of a particle is defined only relative to another system. Or, in the words of their inventor: ``Quantum mechanics is a theory about the physical description of physical systems relative to other systems, and this is a complete description of the world''~\cite{Rovelli1996}. These reference systems may or may not be observers --- the observer hence plays no fundamental role in RQM.

\begin{tcolorbox}
\textbf{Observers in relational quantum mechanics} \\
In relational quantum mechanics, the observer has a status comparable to that in special relativity, namely as a physical system that serves as a reference.  Statements about the outcome of a measurement on a system are always statements about the relation between this reference and the measured system. 
\end{tcolorbox}

Applied to the Wigner's friend thought experiment, RQM yields two different descriptions, one from Alice's viewpoint and one from Wigner's. Relative to Alice, the measurement has a definitive outcome, either $a=0$ or $a=1$. Relative to Wigner, the situation is described by a superposition state of the form~\eqref{eq_WignerFriendState}. That is, for Wigner, there is full symmetry between $a=0$ and $a=1$. RQM endorses such differing viewpoints --- ``different observers can give different accounts of the same set of events''~\cite{Rovelli1996}. Note that this is similar to neo-Copenhagen, where the individual agents' accounts may differ, too.

\subsection{QBism}
\label{subsec:qbism}

\emph{QBism} is an interpretation of quantum mechanics which puts the agent performing a measurement into the spotlight of the theory~\cite{Caves2002, Fuchs2014, Fuchs2016,StanfordQuantum-bayesian}. QBism holds that a quantum state, rather than representing a physical fact, must be regarded as the epistemic state of an agent who reasons about the world around them. It describes the agent's expected experiences (measurement outcomes), depending on the agent's actions (choices of measurements). Concretely, the quantum state is taken to be a concise representation of probabilities assigned to all possible future measurement outcomes. Each of these probabilities quantifies the agent's degree of belief that the corresponding outcome will occur. 

QBist observers have a similar role as agents in Bayesian probability theory~\cite{Cox1946, Jaynes1970}. The probabilities they assign to the possible future observations are subjective, i.e., different agents may make different statements about the same experiment. Similarly to neo-Copenhagen and RQM, these differing viewpoints are a genuine feature of the approach  --- QBism does not impose any compatibility requirements for the beliefs or knowledge of different agents (see however the discussion in~\cite{deBrota2020}).

\begin{tcolorbox}
\textbf{Observers in QBism} \\
According to QBism, quantum theory is a normative theory that is used by individual observers to make decisions. Quantum states are thus subjective: they represent the knowledge or beliefs an observer has about the world around them, rather than an objective fact about the world.  
\end{tcolorbox}

In the Wigner's friend thought experiment, Alice knows after her measurement whether $a=0$ or $a=1$, and she would include this knowledge into the state she assigns to the system. Wigner, as an outside agent, doesn't have that knowledge and hence assigns a different state to the system. His state reflects his belief that, if he asked Alice about the outcome, the two answers $a=0$ and $a=1$ would be equally plausible. Alice and Wigner's state assignments are thus different --- but according to QBism this is unproblematic and just reflects the fact that different agents generally have different knowledge.

\subsection{ETH approach}
\label{subsec:eth}

The starting point of the \emph{ETH approach}~\cite{Froehlich2015} is an algebra of observables. The algebra specifies the set of possible measurements that can in principle be carried out at a given time. The choice of the algebra  corresponds to the choice of the Heisenberg cut in the Copenhagen interpretation. A system is considered \emph{classical} if there does not exist an observable in the algebra that allows to distinguish a superposition of different configurations of the system from a mixture. 

Applied to the Wigner's friend scenario, one has a choice between treating Alice as a classical or as a quantum system, and this choice is made by selecting a suitable algebra. If the algebra contains observables that could confirm the existence of a superposition of $a=0$ and $a=1$ after Alice's measurement of the system, then Alice cannot be regarded as a classical system.  Consequently, the measurement she carried out is not a legitimate measurement according to the ETH interpretation, and  the statement that she observed $a=0$ or $a=1$ is disallowed. Conversely, if the algebra doesn't include any observables that can test for a superposition of $a=0$ and $a=1$ then Alice's measurement is legitimate and has a definite outcome. But since the algebra doesn't allow to distinguish a superposition state like~\eqref{eq_WignerFriendState} from a classical mixture, claiming that Alice is in a superposition state is now vacuous.  

Hence, depending on the algebra of observables one considers, either Wigner can  claim that Alice is in a superposition state, or Alice observes a definitive outcome, but not both. The ETH approach itself however does not provide a guideline as to how to choose the algebra. In any case, as long as one sticks to one choice, the contradiction hinted at by Wigner is avoided. 

\subsection{Consistent histories}
\label{subsec:consistent}

In the \emph{consistent histories}~\cite{Griffiths1984,Omnes1988, StanfordConsistent-histories} (also known as \emph{decoherent histories}~\cite{Gellmann1997}) approach  to quantum mechanics, quantum dynamics corresponds to a stochastic process that produces a sequence of events, called \emph{histories}. The act of measurement (and, hence, the observer) does not play any special role. Rather, to analyse an experiment within this approach, one has to choose a set of possible histories, called a \emph{framework}. The consistent histories interpretation then provides rules to assign probabilities to the individual histories of this framework. Different frameworks can in principle be incompatible, i.e., the predictions the theory makes for one choice of a framework does not need to agree with the predictions obtained within another framework. 

The consistent histories interpretation does not provide any guidance on how to choose such a framework. It can thus, similarly to the ETH approach, be applied in different ways to the Wigner's friend thought experiment; the statements one arrives at depend on the framework one chooses. For one of the choices it is legitimate to claim that Alice's act of measuring leads to a superposition state of the form~\eqref{eq_WignerFriendState}. For another choice of the framework, Alice's measurement outcomes $a=0$ and $a=1$ are legitimate events. The two frameworks cannot in general be combined into one, though. It is thus up to the user of the theory to decide which of the statements are more physically accurate~\cite{Griffiths1998}.

\subsection{Objective collapse theories}
\label{subsec:collapse}

\emph{Objective collapse theories} do not offer an interpretation, but rather a modification of quantum theory~\cite{GisinStochastic,Ghirardi1985,StanfordCollapse}.  Various versions of collapse theories have been proposed in the literature, and they differ by the nature of the modification. Their common feature is that they introduce a mechanism that causes superposition states in  macroscopic systems to decohere on very short timescales. Macroscopic systems may therefore be treated as classical systems. At the same time, this decoherence mechanism is designed in such a way that its effect on microscopic quantum systems, for which standard quantum theory has been confirmed experimentally, is negligible.\footnote{With more and more refined experiments on larger and larger systems it has been possible to narrow down the range of possible collapse models.} The decoherence mechanism thus yields a separation into quantum and classical behaviour, and in this sense implies a Heisenberg cut which is observer-independent and hence objective. 

One of the most prominent concrete collapse theories is the \emph{Ghirardi-Rimini-Weber (GRW) model}~\cite{Ghirardi1985, Ghirardi1986}. It postulates that each elementary  physical system, such as a particle, is subject to random and spontaneous localisation processes. They are captured by nonlinear terms that are added to the standard quantum-theoretic equation of motion, i.e., the Schr{\"o}dinger's equation. These terms depend on phenomenological parameters, for example, the localization accuracy and the localization frequency, which govern how often spontaneous localization processes occur and how they look like. Of a similar kind is the \emph{Di\'osi-Penrose model}~\cite{Diosi1989,Penrose1999}. It postulates a mechanism that causes superposition states whose components have large energy differences to decohere quickly.

Collapse theories do not assign a special role to observers. Moreover, the very purpose of collapse models is to justify the classicality of observers. Hence, when analysing the Wigner's friend thought experiment within a collapse theory, it is natural to treat Alice as a classical system. Superposition states of the form~\eqref{eq_WignerFriendState} are then immediately ruled out.

\begin{table}[]
\centering
\scalebox{0.7}{
\begin{tabular}{|l|l|l|}
\hline
\multicolumn{3}{|c|}{\textbf{Observer required}}     \\ \hline
\textit{Interpretation }             & \textit{Features}   & \textit{Range of applicability of QT} \\ \hline
Conventional Copenhagen  & objective cut between the observer and the observed  & any system under the cut       \\ \hline
Neo-Copenhagen              & subjective cut between the observer and the observed   & any system under the cut      \\ \hline
QBism                       & theory applied from perspective of observer    & any system excluding the observer    \\ \hline
Many-worlds     & measurements by observer induce branching into worlds   & entire universe      \\ \hline\hline
\multicolumn{3}{|c|}{\textbf{No observer required}}    \\ \hline
\textit{Interpretation }                   & \textit{Features} & \textit{Range of applicability of QT} \\ \hline
Bohmian mechanics               & complements quantum theory with hidden variables  & entire universe     \\ \hline
Relational quantum mechanics                     & description always relative to another system & any system in relation to another \\ \hline
ETH approach                      & considers restricted set of observables &  dependent on set of observables    \\ \hline
Consistent histories        & considers restricted set of possible events    & dependent on set of events      \\ \hline
Objective collapse theories (GRW) & modification of Schr\"odinger equation yields non-unitarity   & microscopic systems     \\ \hline
Montevideo interpretation         & gravitation induces non-unitarity   & microscopic systems     \\ \hline \hline
\end{tabular}}
\caption{{\bf The role of observers in different interpretations.} Interpretations of quantum theory can be categorised by how they treat observers. The upper half of the table lists some of the most prominent interpretations that require the notion of an observer. The lower half gives examples of interpretations in which the observer plays no fundamental role. The various interpretations also differ by what they consider the range of applicability of quantum theory.}
\label{tab:observer}
\end{table}

\subsection{Montevideo interpretation}
\label{subsec:montevideo}

The \emph{Montevideo interpretation} postulates that classical gravitational effects are relevant for the quantum measurement process~\cite{Gambini2015}.
They place limitations on the accuracy to which we can determine physical quantities, in particular how precisely we can measure time. These limitations are regarded as fundamental, which makes time  an intrinsically uncertain parameter. Taking into account this intrinsic uncertainty, the time evolution of a quantum system becomes non-unitary. Like in collapse theories, this implies that large systems behave classically. 

\begin{tcolorbox}
\textbf{Observers in the ETH approach, objective collapse theories, consistent histories, and the Montevideo interpretation} \\
According to these approaches the notion of an observer does not have a role in quantum theory, i.e., none of the statements about an experiment that these interpretations endorse depends on an observer.
\end{tcolorbox}

Like collapse theories, the Montevideo interpretation imposes a size limit on systems which can be in a superposition state. For experiments that include agents, like Wigner's friend, this means that superposition states of the form~\eqref{eq_WignerFriendState} cannot occur.

\subsection{Minimal characterisation of an observer} 

Physicists have not reached a consensus on how quantum theory should be interpreted. The role that the observer plays in quantum theory, if any at all, is thus still debated. This is not usually a problem when applying quantum theory to current experiments, where the systems under investigation are microscopic. Here all interpretations agree on their predictions for the experimental observations. Ambiguities arise however if quantum systems are complex enough so that they can contain observers. The purpose of this review is precisely to discuss such situations. We should thus clarify what we mean by an \emph{observer}. But since there is no generally agreed-upon definition for this concept, we need a characterisation that is as minimal as possible (and thus compatible with most interpretations), yet sufficient to capture the features that are relevant to our discussion. 

We regard an \emph{observer} or an \emph{agent} simply as a user of the theory, i.e., a system that is able to carry out physics experiments and apply the laws of quantum theory to make predictions (or retrodictions) about them. An agent could thus be a human observer or just a simple machine that executes certain operations following a sequence of instructions. These instructions may include measuring systems, storing information about the outcomes in an internal memory, processing this information and drawing conclusions by logical reasoning within the set of rules specified by quantum theory, and issuing statements about future (or past) measurement outcomes. In the Wigner's friend thought experiment, the instructions Alice must follow are quite straightforward. She first has to measure the system~$R$ and then state her conclusion that $a=0$ or that $a=1$.



\section{Wigner's friend-in-a-box thought experiment}
\label{sec:wigner}
Devising thought experiments is a long-running tradition in science in general, and in physics in particular. Thought experiments are used for different purposes, from pure entertainment to conceptual analysis and theory testing, where they can be a decisive tool~\cite{TEStanford}. They usually point to a surprising conclusion or an apparent contradiction that arises when  a physical theory is applied to specific scenarios. Among the well-known examples are the Maxwell's demon thought experiment, which we mentioned in the introduction. Another prominent example from thermodynamics is the Gibbs paradox~\cite{Gibbs1875, Jaynes1992}, where two identical containers of an ideal gas are mixed and then separated again. This thought experiment examines the notions of entropy and indistinguishability of particles. Further examples include the famous twin paradox in special relativity~\cite{Einstein1911}, illustrating the effect of time dilation, and Norton's dome thought experiment~\cite{Norton2003}, showing that Newtonian mechanics can exhibit non-deterministic behaviour. Quantum thought experiments often focus on the measurement process. This is in particular true for the Wigner's friend experiment as well as its extensions, which we are now going to discuss in more detail.


\subsection{Description of the thought experiment}
\label{subsec:wigner-description}

\begin{figure}[p]
\centering
\includegraphics[scale=0.25]{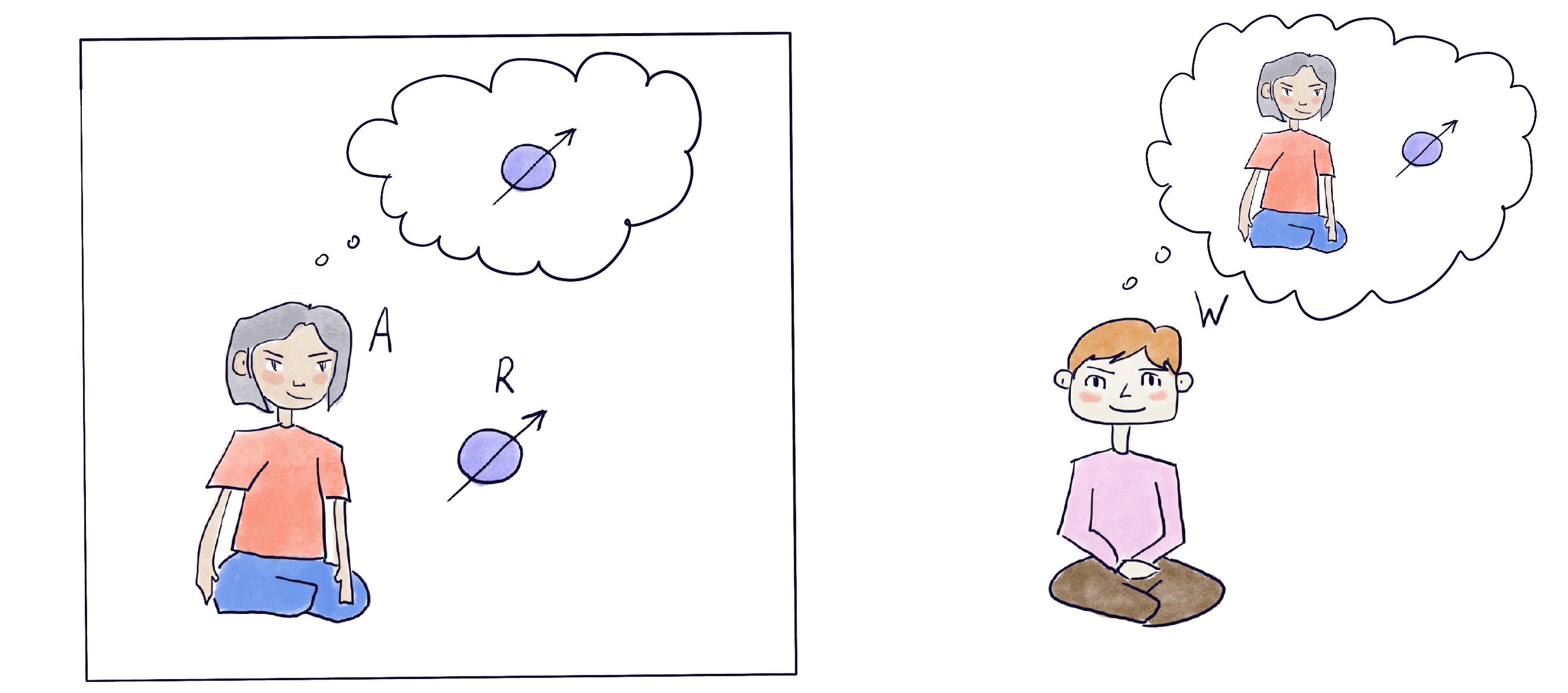}
\caption{{\bf Wigner's friend-in-a-box experiment.} Alice carries out her measurement of system $R$ in an isolated box or lab. For Wigner, the lab may thus be treated as a big quantum system that undergoes a unitary time evolution.}
\label{fig:wigner}
\end{figure}

The starting point of our discussion is a variant of the Wigner's friend thought experiment, where an agent Alice is located in a box or a lab that is perfectly isolated (Figure~\ref{fig:wigner}). Like before, Alice measures a system $R$, initialised to state~\eqref{eq_SystemInitial}, with respect to the computational basis $\{\ket{0}_R,\ket{1}_R\}$, and records the outcome, $a=0$ or $a=1$. While doing this, she remains in her isolated lab. Meanwhile, another agent, Wigner, describes this experiment from the outside. He treats Alice and her lab as a physical system, and we assume that he has full knowledge about this system. Furthermore, from Wigner's perspective, Alice and her lab shall initially be in a pure state. 

A crucial assumption underlying the Wigner's friend thought experiment is that Alice's entire lab, including Alice herself, is subject to the laws of quantum theory. In particular, Alice's memory, in which she stores her observations and conclusions, is a quantum system, which we denote by~$A$, and which we assume to be equipped with a discrete set of basis states. It is useful to label them by statements that Alice may make, such as $\states{I observed $a=0$.}{A}$ or $\states{I observed $a=1$.}{A}$. While this abstraction captures all that is relevant for our considerations, we emphasise that Alice may of course be an arbitrarily complex physical system. 

Although we have already done this briefly in the introduction, it is worth reminding us of how the experiment looks like from Wigner's perspective. Wigner knows that Alice's memory is initially in a pure state, which we may call $\states{I am ready.}{A}$. Let us denote by $\bar{A}$ the environment within Alice's lab, i.e., everything in Alice's lab except for $R$ and Alice's memory~$A$. Alice's lab thus forms a composite system consisting of  subsystems   $R$, $A$, and $\bar{A}$. The state transition during Alice's measurement and recording of the result is thus of the form
\begin{align} 
   \ket{0}_R \otimes \states{I am ready.}{A}  \otimes \ket{\text{env}}_{\bar{A}}
   & \quad \longmapsto \quad
   \ket{\text{lab}_0}_{RA\bar{A}}:=\ket{0}_R \otimes \states{I observed $a=0$.}{A}  \otimes \ket{\text{env}_0}_{\bar{A}} \nonumber \\
    \ket{1}_R \otimes \states{I am ready.}{A}  \otimes \ket{\text{env}}_{\bar{A}} 
   & \quad \longmapsto \quad
   \ket{\text{lab}_1}_{RA\bar{A}} := \ket{1}_R \otimes \states{I observed $a=1$.}{A}  \otimes \ket{\text{env}_1}_{\bar{A}} \label{eq_AliceMeasurement}  \ ,
\end{align}
where $\ket{\text{env}}_{\bar{A}}$ denotes the initial pure state of the environment within Alice's lab, and where $\ket{\text{env}_a}_{\bar{A}}$ is again a pure state of the environment that may depend on $a \in \{0,1\}$. Since by assumption the lab is isolated, its time evolution must be unitary. Wigner knows that $R$ is initialised to a superposition state. He must thus conclude, by linearity, that Alice's lab ends up in the superposition state 
\begin{align} \label{eq_AliceSuperposition}
\ket{\text{lab}_+}_{R A \bar{A}} = \sqrt{\frac{1}{2}}\bigl(\ket{\text{lab}_0}_{RA\bar{A}}
    +\ket{\text{lab}_1}_{RA\bar{A}} \bigr) \ .
\end{align}
This of course corresponds to~\eqref{eq_WignerFriendState}, except that we now assumed that Alice is in an isolated lab, so that it suffices to consider the environment within it, which we labelled~$\bar{A}$.

\begin{figure}[p]
    \begin{subfigure}[t]{0.9\textwidth}
    \centering
        \includegraphics[scale=.45]{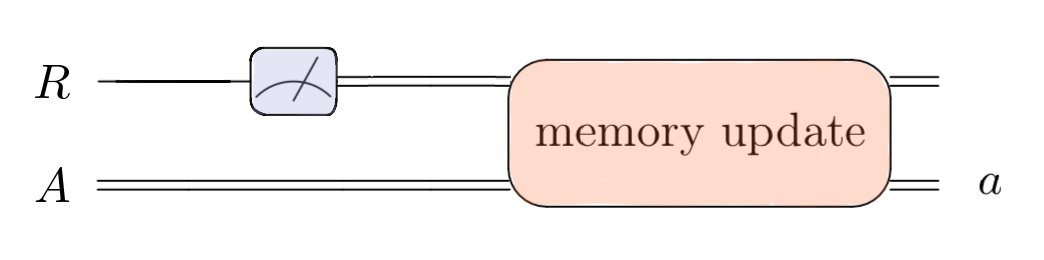}
        \caption{Alice's view: the quantum system $R$ is measured and the result is written into the memory~$A$, which is treated as a classical system.}
        \label{fig:wignerA}
    \end{subfigure}
    \\
    \begin{subfigure}[t]{0.9\textwidth}
    \centering
        \includegraphics[scale=.45]{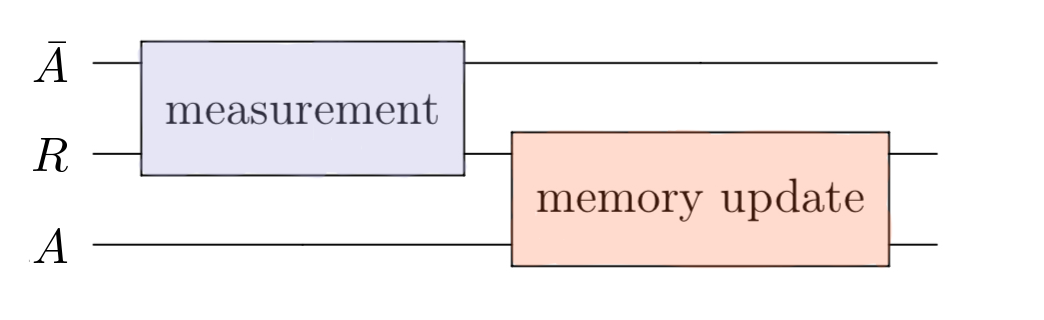}
        \caption{Wigner's view:  The measurement on $R$ carried out by Alice may also influence degrees of freedom in Alice's environment~$\bar{A}$ within her lab. All systems within Alice's lab, $R$, $A$, and $\bar{A}$, are treated as quantum systems, and since the lab is assumed to be isolated, their evolution is unitary.}
        \label{fig:wignerW}
    \end{subfigure}
    
\caption{\textbf{Circuit representation of agents' views on the Wigner's friend experiment.} Each circuit shows the view of one agent on the experiment. Single wires represent systems that are treated as quantum systems, and double wires stand for systems that are regarded as classical, in the sense that they have a definite value for the agent. For example, Alice's memory $A$ is classical for Alice but quantum for Wigner. The boxes (or gates) represent individual operations carried out by the agents. Boxes with identical colour correspond to the same operation. Depending on the agent's view, an operation may be unitary or not. For example, for Alice the measurement of $R$ is a non-unitary operation (indicated by round corners) acting on $R$ only, whereas for Wigner it is a unitary (rectangular corners) acting on $R$ and Alice's environment $\bar{A}$.}
\label{fig:wignercirc}
\end{figure}

Alice and Wigner's view on the thought experiment may be represented by circuit diagrams as shown in Figure~\ref{fig:wignercirc}. The two diagrams both refer to the same experiment, but their differences reflect the different perspectives that two agents have on it.  For example, Alice would typically only describe $R$ as a quantum system, but not herself or her environment $\bar{A}$. Furthermore, she would treat her own memory as a classical register. This is shown in Figure~\ref{fig:wignerA}. Conversely, Wigner views $R$, $A$, and $\bar{A}$ as quantum systems, as depicted by Figure~\ref{fig:wignerW}. Each single wire in the circuit corresponds to a quantum subsystem, and a double wire to a system that is regarded as classical. Each box corresponds to an operation (also called a \emph{gate}) carried out by an agent, and these are applied from left to right.

\subsection{Implications}

As discussed in Section~\ref{sec:wigner}, different interpretations of quantum theory yield different statements about the Wigner's friend experiment. Let us here briefly summarise how they answer one of the central questions raised by the thought experiment: What does the superposition state~$\ket{\text{lab}_+}_{R A \bar{A}}$ defined by~\eqref{eq_AliceSuperposition} mean, and why does it not contradict the fact that Alice perceives a definitive outcome, either $a=0$ or $a=1$? Wigner's own conclusion was that the state assignment~\eqref{eq_AliceSuperposition} must be wrong, and that standard quantum theory is therefore not applicable to describe (conscious) observers. This is however only one of many possible ways out of the apparent contradiction. 
\begin{itemize}
    \item According to collapse interpretations, Alice is a macroscopic system that is subject to fundamental decoherence. Her lab would thus, rather than remaining in the superposition state~\eqref{eq_AliceSuperposition}, very quickly evolve towards a mixture of the two states $\ket{\text{lab}_0}_{R A \bar{A}}$ and  $\ket{\text{lab}_1}_{R A \bar{A}}$ defined by~\eqref{eq_AliceMeasurement}, each of which describes a situation where Alice observed a definitive value, $a=0$ or $a=1$. 
    \item According to interpretations that postulate an objective Heisenberg cut (such as conventional Copenhagen, as well as the ETH approach and consistent histories\footnote{The latter do not explicitly postulate such a cut. However, the choice of an observable algebra in the ETH approach or the choice of a framework in the consistent histories approach implies an effective quantum-classical cut.}), the cut decides whether Alice is a classical or a quantum system. In the first case, assigning a quantum state like $\ket{\text{lab}_+}_{R A \bar{A}}$ to Alice's lab is disallowed. In the second case, where Alice is a quantum system, she cannot be an observer, and hence saying that she observed $a=0$ or $a=1$ has no meaning. 
    \item According to interpretations that postulate a subjective Heisenberg cut (such as neo-Copenhagen, QBism, or relational quantum mechanics), statements are always relative to an agent and should not be compared. It is hence correct to say that, from Alice's viewpoint, a definitive outcome $a=0$ or $a=1$ was observed. At the same time it is also correct to say that, from Wigner's perspective, Alice's lab is in the superposition state~$\ket{\text{lab}_+}_{R A \bar{A}}$. 
    \item According to interpretations that postulate a universal wave function of the universe (such as Bohmian mechanics or Everett's many-worlds interpretation),  Alice's lab is in the superposition state~$\ket{\text{lab}_+}_{R A \bar{A}}$, but  this does not contradict the statement that Alice observes a definitive outcome, $a=0$ or $a=1$. In the case of Bohmian mechanics, the actual outcome is determined by classical hidden variables. In the case of many-worlds, both possible observations of Alice are equally real.    
\end{itemize}

One may conclude that all these interpretations manage to somehow circumvent a contradiction between Alice's and Wigner's viewpoints. However, the situation becomes more intricate as we consider extensions of this thought experiment.

\section{Wigner-Deutsch thought experiment}
\label{sec:deutsch}
Two central questions that came up repeatedly in our discussions so far are  (i)~``Does Alice see a definitive measurement outcome?'' and (ii)~``Is Alice's lab after the measurement indeed in a superposition state?'' Deutsch~\cite{Deutsch1985} proposed an extension of the thought experiment described in Section~\ref{sec:wigner}, which turns these questions into (in principle) experimentally testable statements.

\subsection{Description of the thought experiment}
\label{subsec:deutsch-description}

\begin{figure}[p]
\centering
\includegraphics[scale=0.25]{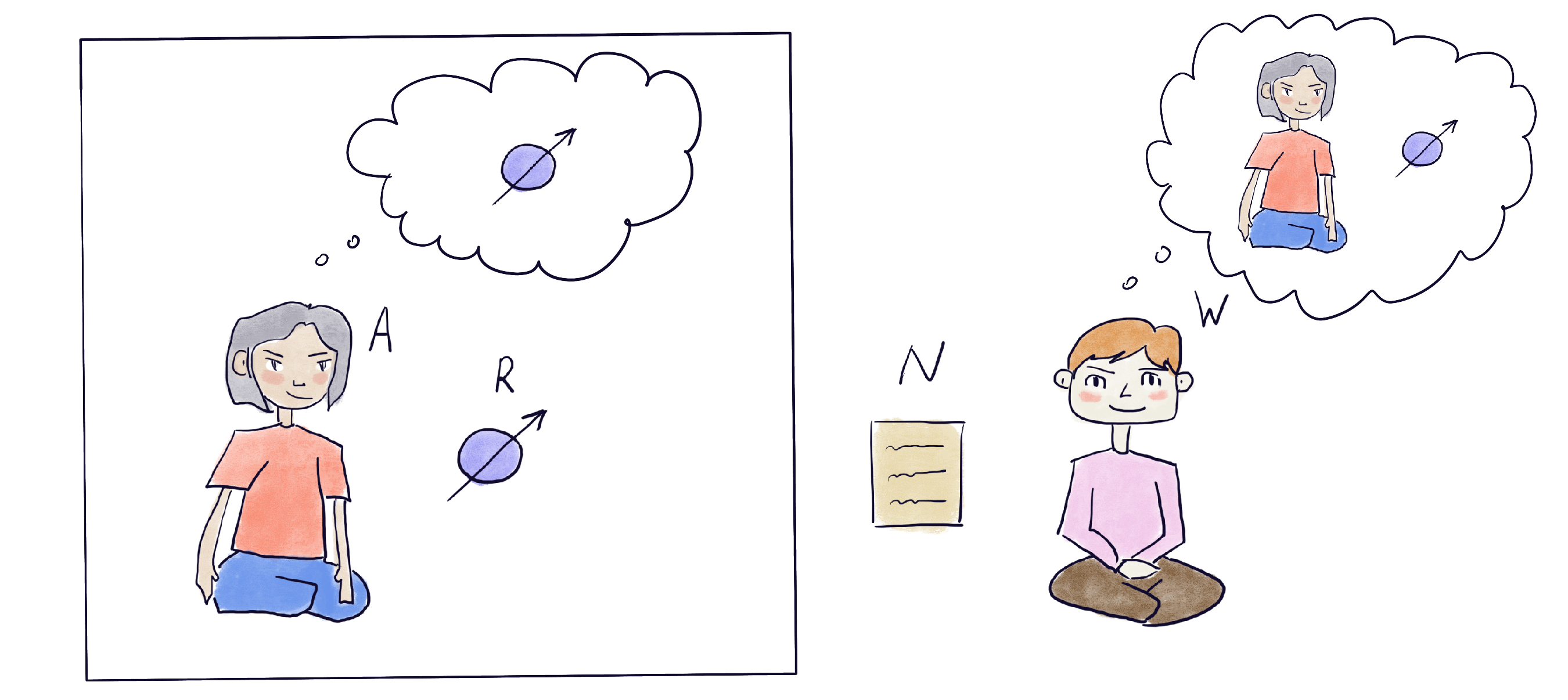}
\caption{{\bf Wigner-Deutsch thought experiment.} Deutsch's extension to the thought experiment of Figure~\ref{fig:wigner} features an extra system $N$, a notebook in which  Alice writes whether or not she has observed a definite measurement outcome. Wigner additionally carries out a measurement on Alice's lab  (not shown in the picture) to check whether it is in a superposition state.}
\label{fig:deutsch}
\end{figure}

To address question~(i), Deutsch considers some (limited) communication between Alice and Wigner. Specifically, Deutsch equips Alice with a \emph{notebook}~$N$, in which she writes whether or not she observed a definitive measurement outcome, but does not reveal the outcome itself. This notebook is then handed to Wigner (Figure~\ref{fig:deutsch}). The resulting state transition of Alice's lab remains basically the same as~\eqref{eq_AliceMeasurement}, except that it now also includes~$N$, i.e.,
\begin{align} 
       \ket{0}_R\otimes\states{I am ready.}{A}  \otimes \ket{\text{env}}_{\bar{A}} \otimes \ket{\text{empty}}_N & \quad \longmapsto \quad \ket{\text{lab}_0}_{R A \bar{A}} \otimes \states{I observed a definitive outcome.}{N}  \nonumber \\
         \ket{1}_R\otimes\states{I am ready.}{A}  \otimes \ket{\text{env}}_{\bar{A}} \otimes \ket{\text{empty}}_N& \quad \longmapsto \quad
       \ket{\text{lab}_1}_{R A \bar{A}} \otimes \states{I observed a definitive outcome.}{N}  \label{eq_EvolutionDeutsch} \ .
\end{align}
By linearity, the joint state of Alice's lab and the notebook will thus be
\begin{align} \label{eq_DeutschNotebook}
   \underbrace{\sqrt{\frac{1}{2}}\bigl(\ket{\text{lab}_0}_{RA\bar{A}}
    +\ket{\text{lab}_1}_{RA\bar{A}} \bigr)}_{\ket{\text{lab}_+}_{R A \bar{A}}} \otimes \states{I observed a definitive outcome.}{N}  \, 
\end{align}
i.e., Alice's lab is again in the superposition state~\eqref{eq_AliceSuperposition} we encountered in the previous version of the experiment. However, Wigner now gets a certificate from Alice --- a note written down by her confirming that she indeed observed a definitive outcome. It is crucial though that the note does not contain any information about the actually observed value, $a=0$ or $a=1$. If it did, Alice's lab and the notebook would no longer be in a product state as in~\eqref{eq_DeutschNotebook}, but would be entangled instead. Furthermore, if Wigner could read, say, that $a=0$, he would conclude that Alice's lab is now in state $\ket{\text{lab}_0}_{R A \bar{A}}$ rather than in the superposition state~$\ket{\text{lab}_+}_{R A \bar{A}}$.

To answer question~(ii), Deutsch added  a final step to the experiment, where Alice's measurement of $R$ is \emph{undone}. That is, after Alice has completed her measurement and sent the notebook to Wigner, some external control is applied to Alice's entire lab that induces the state transition inverse to~\eqref{eq_EvolutionDeutsch}, 
\begin{align}
      \ket{\text{lab}_0}_{R A \bar{A}} 
& \quad \longmapsto \quad
       \ket{0}_R\otimes \states{I am ready.}{A} \otimes  \ket{\text{env}}_{\bar{A}}  \nonumber
       \\
       \ket{\text{lab}_1}_{R A \bar{A}} 
       & \quad \longmapsto \quad
         \ket{1}_R\otimes\states{I am ready.}{A} \otimes \ket{\text{env}}_{\bar{A}} \label{eq_MeasurementReverse}
        \ .
\end{align}
Such a reverse evolution of Alice's entire lab may be difficult to realise (and be practically impossible if Alice was a human observer), but not fundamentally disallowed by quantum theory. Thus assuming that it can be implemented, the resulting state of Alice's lab is 
\begin{align} \label{eq_RevertedLab}
   \ket{+}_R \otimes \states{I am ready.}{A} \otimes  \ket{\text{env}}_{\bar{A}} \ ,
\end{align}
i.e., Alice's record of the measurement outcome is erased and the system $R$ returned to its initial superposition state $\ket{+}_R$ defined by~\eqref{eq_SystemInitial}. Finally, the outside agent Wigner may test this superposition by applying a direct measurement to~$R$ that includes a projection onto~$\ket{+}_R$. According to quantum theory, the projection must succeed with probability~$1$. 

Conversely, if the state of Alice's lab, after she completed her measurement of $R$, had collapsed to $\ket{\text{lab}_0}_{R A \bar{A}}$ or $\ket{\text{lab}_1}_{R A \bar{A}}$, the state of $R$ after the undoing operation~\eqref{eq_MeasurementReverse} would be either $\ket{0}_R$ or $\ket{1}_R$, rather than $\ket{+}_R$ as in~\eqref{eq_RevertedLab}. Hence, in this case, Wigner's final measurement with respect to~$\ket{+}_R$ would have a random outcome. Wigner's measurement can thus distinguish whether or not Alice's lab was in a superposition state and with that answer question~(ii).

\begin{figure}[p]
\centering
    \begin{subfigure}{0.9\textwidth}
        \centering
        \includegraphics[scale=.45]{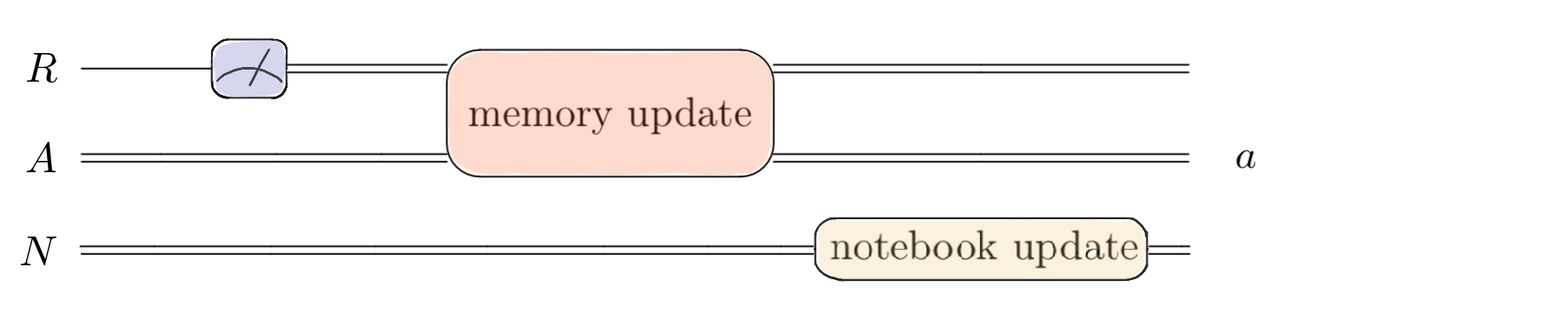}
        \caption{Alice's view: first system $R$ is measured and the result is written down to her memory (which is classical); after the outcome is observed, the state of the notebook is updated.}
        \label{fig:deutschA}
    \end{subfigure}
    \
    \begin{subfigure}{0.9\textwidth}
        \includegraphics[scale=.45]{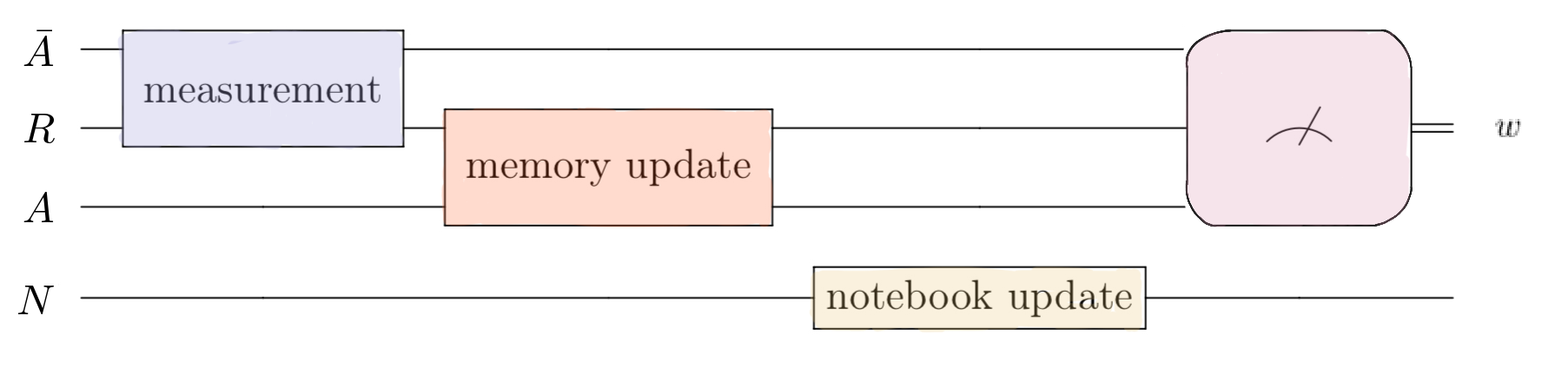}
        \caption{Wigner's view: in her isolated lab, Alice measures the system $R$, writes the result to her memory (which, for Wigner, is a quantum system), and updates the notebook accordingly. The box on the right represents the projective measurement with respect to state $\ket{\text{lab}_+}_{R A \bar{A}}$ which Wigner applies to Alice's lab.}
        \label{fig:deutschW}
    \end{subfigure}
    
\caption{\textbf{Circuit representation of agents' views on the Wigner-Deutsch  experiment.} The diagrammatic language and colour code is the same as before (see Figure~\ref{fig:wignercirc}). An agent may restrict their quantum-theoretic description to those parts of the experiment that are relevant for their own actions.  For example, according to quantum theory, the final measurement that Wigner applies to Alice's lab (pink box) does not have an impact on the earlier operations carried out by Alice, and Alice may thus omit it from her description.}
\label{fig:deutschcirc}
\end{figure}

Like Wigner's original thought experiment, we may represent Deutsch's variant by an appropriately extended circuit diagram (Figure~\ref{fig:deutschcirc}). From Alice's perspective, writing into the notebook is a classical operation that does not depend on the actual outcome of the measurement. The idea that the notebook entry depends on whether or not she observed a definitive outcome, however, cannot be reflected by the circuit diagram. The reason is that  circuit diagrams are just a graphical way of representing elements of the standard quantum formalism. And, according to this formalism, a measurement on a quantum system always produces an outcome. 

The undoing operation~\eqref{eq_MeasurementReverse} followed by the measurement  of $R$ with respect to $\ket{+}_R$ produces the same outcome as a direct  measurement of  Alice's lab with respect to the state $\ket{\text{lab}_+}_{R A \bar{A}}$ defined by~\eqref{eq_AliceSuperposition}. We may thus use the latter to keep the circuit diagram representation as simple as possible. Note that we are not interested in the state of Alice's lab after this measurement, but only in the classical measurement outcome, which we denote by $w$.

\subsection{Implications}
\label{subsec:deutsch-assumptions}

Like for Wigner's original thought experiment, different interpretations of quantum theory lead to different conclusions about Deutsch's extension, which we summarise here.
\begin{itemize}
   \item According to collapse interpretations, Alice's lab would be subject to decoherence and thus evolve towards a probabilistic mixture of $\ket{\text{lab}_0}_{R A \bar{A}}$ and $\ket{\text{lab}_1}_{R A \bar{A}}$. Wigner's final measurement of Alice's lab with respect to $\ket{\text{lab}_+}_{R A \bar{A}}$ would thus yield a random outcome, indicating that Alice's lab was not in that superposition state.
   \item According to interpretations that postulate an objective Heisenberg cut, if $A$ is above the cut and thus a classical system, the final measurement of Alice's lab as described above cannot be carried out. In fact, the measurement is ill-defined, as it involves the projection onto a superposition state, namely $\ket{\text{lab}_+}_{R A \bar{A}}$, of a partially classical system. Conversely, if $A$ is under the cut and thus a quantum system, this final measurement is well-defined and projects with probability~$1$ onto $\ket{\text{lab}_+}_{R A \bar{A}}$. However, in this case  Alice cannot be treated as an observer, and the statement she is supposed to write into the notebook, that she observed a definitive outcome, does not make sense. 
   \item According to interpretations that postulate a subjective Heisenberg cut, Alice would be right to claim that she observed a definitive outcome. Nonetheless, from Wigner's viewpoint, Alice's lab can be a quantum system. Wigner's final measurement of Alice's lab will thus project onto the superposition state $\ket{\text{lab}_+}_{R A \bar{A}}$ with probability~$1$, confirming that it was indeed in that state. 
   \item According to interpretations that postulate a universal wave function of the universe, the results would be the same as above. Alice would be right to claim that she observed a definitive outcome (although, in many-worlds, there would be two branches, one for $a=0$ and one for $a=1$, in each of which such a claim is made). Still Wigner's  measurement will confirm with probability~$1$ that Alice's lab was in the superposition state $\ket{\text{lab}_+}_{R A \bar{A}}$. 
\end{itemize}

The summary exhibits a remarkable feature of Deutsch's extended thought experiment: The different interpretations make different statements about  observable outcomes, such as the result of Wigner's measurement of Alice's lab. This means, in turn, that running the experiment in real would allow us to rule out some of these interpretations.

\section{FR thought experiment}
\label{sec:fr}
In the last step of the Wigner-Deutsch thought experiment, Wigner measures Alice's entire lab to check whether it is in the superposition state  $\ket{\text{lab}_+}_{R A \bar{A}}$. Measurements generally disturb the measured system. In this case the measured system includes Alice's memory, where she stored her information about the value $a$ that she observed when she measured the system~$R$. Hence, if after completion of the experiment we asked Alice  whether she had observed $a=0$ or $a=1$, we cannot expect to obtain a reliable answer.\footnote{In Deutsch's original description, Wigner's measurement of Alice's lab consists of a first step where the unitary evolution of the lab during Alice's measurement of~$R$ is run backwards, so that everything is reset to the state before she started the measurement (see \eqref{eq_MeasurementReverse} and \eqref{eq_RevertedLab}). Hence, if we now asked Alice what outcome she had observed, she would likely just answer that she hasn't even started with her measurement.} (We will come back to this later; see also Figure~\ref{fig:Hadamard}.)

Conversely, as  we already remarked, if Alice, upon completing her  measurement of $R$, immediately communicated the outcome $a$ to the outside, her lab would no longer be in the superposition state $\ket{\text{lab}_+}_{R A \bar{A}}$, but  rather be either in state $\ket{\text{lab}_0}_{R A \bar{A}}$ or $\ket{\text{lab}_1}_{R A \bar{A}}$ (see~\eqref{eq_DeutschNotebook} and the text just below). Wigner would notice this in his final measurement with respect to $\ket{\text{lab}_+}_{R A \bar{A}}$, which would yield a random outcome.

These considerations exhibit a basic dilemma inherent to Wigner's friend-type  experiments: \emph{Either} we can confirm by an outside measurement that Alice's lab is in a superposition state like $\ket{\text{lab}_+}_{R A \bar{A}}$, which has equal components for $a=0$ and $a=1$. \emph{Or} we can ask Alice and learn whether she observed $a=0$ or $a=1$. But it appears impossible to do both in the same run of the experiment.

A key goal of the \emph{FR thought experiment}~\cite{Frauchiger2018}, which we are going to discuss next, is to overcome this dilemma. The idea is to ``save'' Alice's measurement outcome by transferring (partial) information about it to the outside of her lab, while ensuring at the same time that the lab remains in a superposition state, which can then be confirmed via measurements by outside agents. 

\subsection{Description of the thought experiment}
\label{subsec:fr-description}

\begin{figure}[p]
\centering
\includegraphics[scale=0.25]{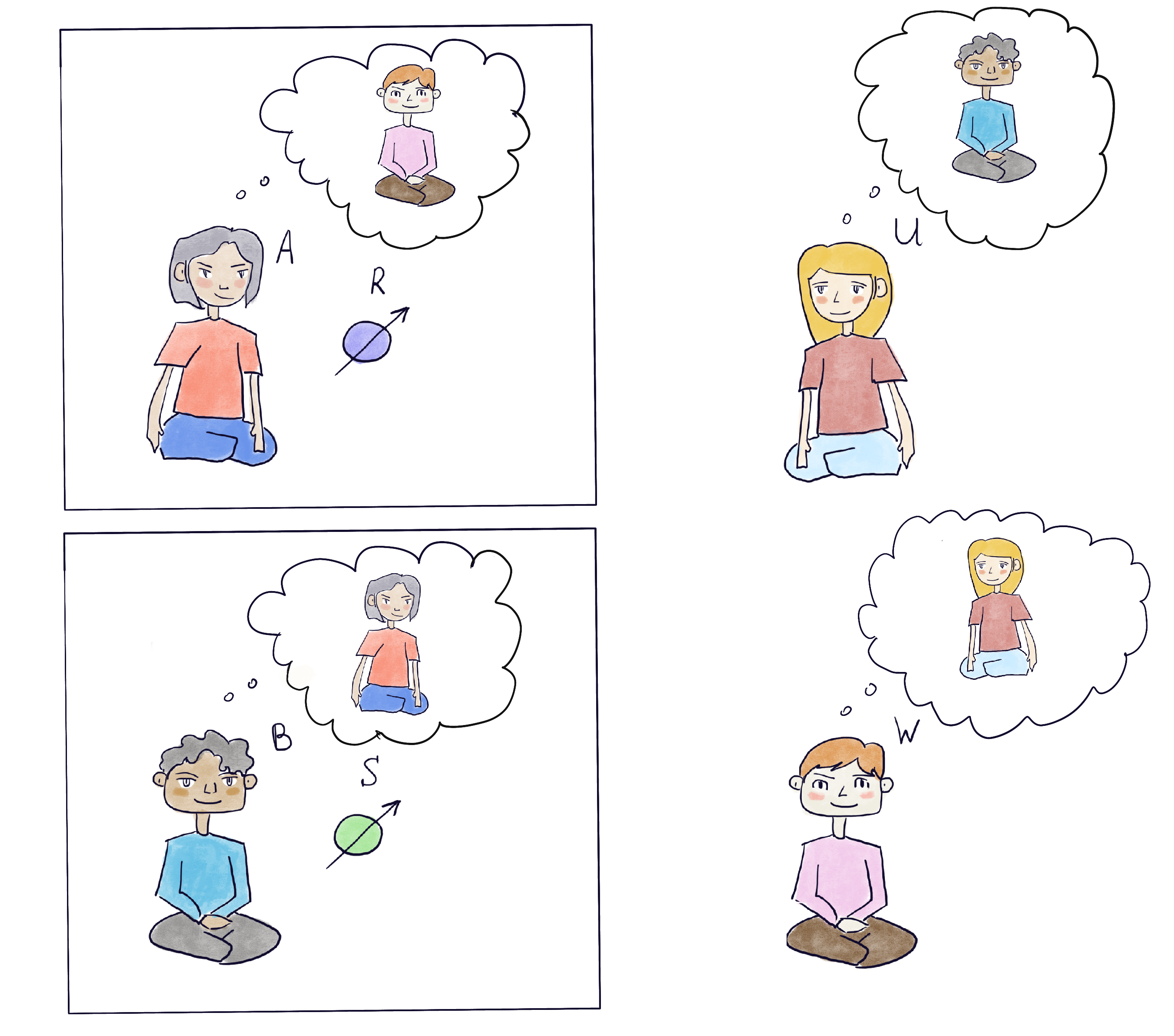}
\caption{\textbf{FR thought experiment.} The experiment features four agents, who are reasoning about each other's observations and predictions. Two of them, Alice and Bob, are each situated in an isolated lab. Alice measures a system $R$ and sends a qubit $S$ to Bob, who measures it. The two other agents, Ursula and Wigner, are located outside of Alice's and Bob's labs and can apply measurements to them.}
\label{fig:fr}
\end{figure}

The FR experiment involves four agents in total (Figure~\ref{fig:fr}). Two of them, Alice and Bob, are located in separate labs, and two agents, Ursula and Wigner, are situated on the outside. Alice measures a system $R$ and is allowed to send one single qubit, $S$, to Bob. Except for this one-qubit communication, Alice and Bob's labs are assumed to be isolated. However, like in the Wigner-Deutsch experiment, we assume that the outside agents can apply measurements to these labs.

A crucial feature of the FR thought experiment  is that the agents are not only performing experimental operations such as measurements, but they are also supposed to make predictions (or retrodictions) about the outcomes of other agents' measurements. For this they apply certain reasoning rules. One of them,  called $\mathbf{Q}$, captures the reasoning process of a user of quantum theory. Another, $\mathbf{C}$, allows an agent to adopt predictions made by another agent. Finally, rule $\mathbf{S}$  reflects the idea that a measurement, at least from the point of view of the measuring agent, has a definite outcome. (The meaning of these rules should already become sufficiently clear from our descriptions below of how the agents apply them; but we will provide more general definitions later in  Section~\ref{subsec:fr-assumptions}.) It is also assumed that the agents possess a full description of the experiment and, in particular, know precisely the tasks performed by the other agents.

The thought experiment is defined by a protocol which specifies the actions  each agent has to carry out. The sequential ordering of the steps is important, and we therefore label each of them by a time~$t$. Like before, the environment in Alice's lab is denoted by~$\bar{A}$. Similarly, $\bar{B}$ stands for the environment in Bob's lab. We assume that Alice's and Bob's memory systems $A$ and $B$, as well as their environments $\bar{A}$ and $\bar{B}$, are initialised to pure states. Finally, we assume that Alice's system $R$ is initialised to  $\smash{\sqrt{\frac{1}{3}}\ket{0}_R+\sqrt{\frac{2}{3}}\ket{1}_R}$. 

\begin{enumerate}[{$t= $} 1:]
    \item Alice measures $R$ in the computational basis $\{\ket{0}_R, \ket{1}_R\}$. Depending on the outcome, she writes either ``I am certain that $a=0$.'' or ``I am certain that $a=1$.'' into her memory~$A$. (This is endorsed by rule~$\mathbf{S}$.) \\
    Alice then prepares the qubit $S$ as follows: if $a=0$ then $S$ is set to $\ket{0}_S$; if $a=1$ then $S$ is set to $\smash{\frac1{\sqrt2}(\ket{0}_S+\ket{1}_S)}$. Once this is done she sends $S$ to Bob. \\
    Alice then uses quantum theory (rule~$\mathbf{Q}$) to predict the outcome, $w$, of the measurement that Wigner will carry out at $t=4$, and writes the prediction in her memory~$A$. (We will explain Alice's reasoning process as well as that of the other agents just below, when we are done with the protocol description.) 

    \item Bob measures $S$ in the computational basis $\{\ket{0}_S, \ket{1}_S\}$ and (again endorsed by rule~$\mathbf{S}$) writes the result in his memory~$B$, e.g.,  ``I am certain that $b=1$.''  \\
    Bob then uses quantum theory (rule~$\mathbf{Q}$) to infer Alice's observation $a$. From this he may in turn infer the prediction Alice came up with for $w$. He adopts this prediction (rule~$\mathbf{C}$) and writes it in his memory~$B$.
    
    \item Ursula measures Alice's lab (consisting of $R$, $A$, and $\bar{A}$) with respect to a basis containing the vectors
    \begin{gather*}
\ket{\text{ok}}_{RA\bar{A}}=\sqrt{\frac{1}{2}}\bigl(\ket{\text{lab}_0}_{R A \bar{A}} - \ket{\text{lab}_1}_{R A \bar{A}} \bigr)\\
\ket{\text{fail}}_{RA\bar{A}}=\sqrt{\frac{1}{2}}\bigl(\ket{\text{lab}_0}_{R A \bar{A}} + \ket{\text{lab}_1}_{R A \bar{A}} \bigr) \ , 
  \end{gather*}
   where $\ket{\text{lab}_0}_{R A \bar{A}}$ and $\ket{\text{lab}_1}_{R A \bar{A}}$ are the states Alice's lab would admit after the step at $t=1$ if $R$ was initialised to $\ket{0}_R$ and $\ket{1}_R$, respectively. She notes down  the outcome, $u=\text{ok}$ or $u=\text{fail}$ (again endorsed by rule~$\mathbf{S}$). \\
   Ursula then uses quantum theory (rule~$\mathbf{Q}$) to infer Bob's observation $b$, from which she may deduce the prediction Bob made for $w$. She  adopts this prediction (rule~$\mathbf{C}$) and announces it publicly. 

    \item Wigner adopts Ursula's publicly announced prediction for $w$ (again rule~$\mathbf{C}$) and notes it down. \\
    Then Wigner measures Bob's lab (consisting of $S$, $B$, and $\bar{B}$) with respect to a basis containing the vectors
    \begin{gather*}
\ket{\text{ok}}_{S B \bar{B}} = \sqrt{\frac{1}{2}}\bigl(\ket{\text{lab}_0}_{S B \bar{B}} - \ket{\text{lab}_1}_{S B \bar{B}} \bigr) \\
\ket{\text{fail}}_{S B \bar{B}} = \sqrt{\frac{1}{2}}\bigl(\ket{\text{lab}_0}_{S B \bar{B}} + \ket{\text{lab}_1}_{S B \bar{B}} \bigr) \ ,
    \end{gather*}
    where $\ket{\text{lab}_0}_{S B \bar{B}}$ and $\ket{\text{lab}_1}_{S B \bar{B}}$ are the states Bob's lab would admit after the step at $t=2$ if $S$ was initialised to $\ket{0}_S$ and $\ket{1}_S$, respectively. He records the observation, which is either $w=\text{fail}$ or $w=\text{ok}$ (rule~$\mathbf{S}$), and compares it with the prediction he noted down just before. 

    \item If the prediction of $w$ is incompatible with the observed value, Wigner declares that the set of reasoning rules must have been contradictory and aborts the experiment. Otherwise he instructs all agents to repeat it. 
\end{enumerate}

Let us now turn to the description of how the individual agents infer their predictions, starting from their records of observations. Note that we only consider particular cases that will be relevant to reach our intended conclusions. 

\begin{itemize}
\item \emph{Alice's inferences, provided she recorded ``I am certain that $a=1$.'':} According to the protocol, she must in this case have prepared $S$ in  state $\smash{\frac{1}{\sqrt{2}}(\ket{0}_S+\ket{1}_S)}$. Hence, by linearity, Bob's state after $t=2$ must be $\smash{\sqrt{\frac{1}{2}}\bigl(\ket{\text{lab}_0}_{S B \bar{B}} + \ket{\text{lab}_1}_{S B \bar{B}} \bigr)}$. But this is precisely one of the basis states, $\ket{\text{fail}}_{SB \bar{B}}$, of Wigner's measurement to obtain~$w$. According to quantum theory ($\mathbf{Q}$), the corresponding outcome must thus occur with certainty. Hence, Alice can  infer the statement ``I am certain that $w=\text{fail}$ at time $t=4$.''

\item \emph{Bob's inferences, provided he recorded ``I am certain that $b=1$.'':}  Suppose by contradiction that Alice observed $a=0$ at time $t=1$. She would then have prepared $S$ in state $\ket{0}_S$, and Bob would, according to quantum theory ($\mathbf{Q}$), with certainty observe $b=0$. But since $b=1$,  Alice must have observed $a=1$. Including now also Alice's reasoning from above, Bob arrives at the  statement ``I am certain that Alice observed $a=1$ and hence she is certain that $w=\text{fail}$ at  $t=4$.'' Bob may now adopt Alice's prediction ($\mathbf{C}$) and conclude ``I am certain that $w=\text{fail}$ at $t=4$.''

\item \emph{Ursula's inferences, provided she recorded ``I am certain that $u=\text{ok}$.'':} The state of Alice's lab at the beginning of the experiment is of the form
\begin{align*}
   \bigl(\sqrt{\frac{1}{3}}\ket{0}_R+\sqrt{\frac{2}{3}}\ket{1}_R\bigr)\otimes \states{I am ready}{A}\otimes \ket{\text{env}}_{\bar{A}} \ .
\end{align*}
After $t=1$ when Alice has completed her task, which includes the preparation of~ $S$, the joint state of her lab and $S$ is by linearity\footnote{Alice's lab and $S$ are now in an entangled state that, up to local transformations, is equivalent to the  state proposed by Hardy in~\cite{Hardy1993} to obtain strengthened versions of Bell's theorem.} 
\begin{align*}
 \sqrt{\frac{1}{3}}\ket{\text{lab}_0}_{RA\bar{A}} \otimes \ket{0}_S+\sqrt{\frac{2}{3}}\ket{\text{lab}_1}_{RA\bar{A}} \otimes \sqrt{\frac{1}{2}} \bigl(\ket{0}_S + \ket{1}_S\bigr) \ .
\end{align*}
For our further considerations, we rewrite this as
\begin{align} \label{eq_JointStateAliceS}
 \sqrt{\frac{2}{3}} 
  \underbrace{\sqrt{\frac{1}{2}}\bigl(\ket{\text{lab}_0}_{RA\bar{A}}+ \ket{\text{lab}_1}_{RA\bar{A}}  \bigr)}_{\ket{\text{fail}}_{RA \bar{A}}} \otimes \ket{0}_{S} +
\sqrt{\frac{1}{3}} \ket{\text{lab}_1}_{RA\bar{A}} \otimes \ket{1}_{S} \ .
\end{align}
One can now readily verify that this sum is orthogonal to the state $\ket{\text{ok}}_{R A \bar{A}} \otimes \ket{0}_S$. Indeed, this is the case for both individual terms of the outer sum of~\eqref{eq_JointStateAliceS}: for the first term, it follows because $\ket{\text{ok}}_{R A \bar{A}}$ is orthogonal to $\ket{\text{fail}}_{R A \bar{A}}$, and for the second because $\ket{0}_S$ is orthogonal to $\ket{1}_S$.  Consequently, according to quantum theory ($\mathbf{Q}$), the probability of the event that Ursula observes $u=\text{ok}$ and Bob observes $b=0$ is zero. Hence, if Ursula did observe $u=\text{ok}$, she can be certain that Bob's outcome was $b=1$. Taking into account Bob's reasoning described above, Ursula may thus note down the statement ``I am certain that Bob observed $b=1$ and hence he is certain that $w = \text{fail}$ at $t=4$.'' Finally, Ursula can adopt Bob's prediction ($\mathbf{C}$) and announce ``I am certain that $w = \text{fail}$ at $t=4$.''

\item \emph{Wigner's inferences, provided Ursula announced ``I am certain that $w=\text{fail}$ at $t=4$.'':} Wigner, after listening to Ursula's announcement and taking into account her reasoning described just above, can issue the statement ``I am certain that Ursula is certain that $w=\text{fail}$ at $t=4$.'' Wigner may now adopt Ursula's prediction ($\mathbf{C}$) and conclude that ``I am certain that $w=\text{fail}$ at $t=4$.''
\end{itemize}

To finalise our analysis, let us consider the event that Ursula and Wigner's measurement outcomes are $u=\text{ok}$ and $w=\text{ok}$, respectively. To calculate the probability of this event, we use~\eqref{eq_JointStateAliceS} as a starting point, which describes the situation before Bob measures $S$. By linearity, the joint state of Alice's and Bob's labs after Bob's measurement at time $t=2$ must be 
\begin{align} \label{eq_JointStateAliceBob}
  \sqrt{\frac{2}{3}} 
\ket{\text{fail}}_{RA \bar{A}} \otimes \ket{\text{lab}_0}_{SB \bar{B}} +
\sqrt{\frac{1}{3}} \ket{\text{lab}_1}_{RA\bar{A}} \otimes \ket{\text{lab}_1}_{SB \bar{B}} \ .
\end{align}
It is straightforward to verify that the overlap between this state and the state $\ket{\text{ok}}_{R A \bar{A}} \otimes \ket{\text{ok}}_{S B \bar{B}}$ equals $\smash{\sqrt{\frac{1}{12}}}$. The probability of the event  $u=w=\text{ok}$ is thus $\frac{1}{12}$. This means that, if the experiment is repeated, the event $u=w=\text{ok}$ will occur at some point. So let's see what happens in this case. 

As we have shown above, if $u=\text{ok}$ then Ursula  announces  ``I am certain that $w=\text{fail}$ at $t=4$.'', in which case Wigner also concludes that ``I am certain that $w=\text{fail}$ at $t=4$.'' Nonetheless, he actually observes  $w=\text{ok}$! He is thus forced to issue the statement 
\begin{align} \label{eq_WignerConclusion}
  \text{``I am certain that $w=\text{fail}$ and I am certain that $w=\text{ok}$.''}
\end{align}
and, consequently, admit that the reasoning rules $\mathbf{Q}$, $\mathbf{C}$, and $\mathbf{S}$ led to a contradiction.  

\begin{figure}[p]
\centering
    \begin{subfigure}{0.9\textwidth}
    \centering
     \includegraphics[scale=.45]{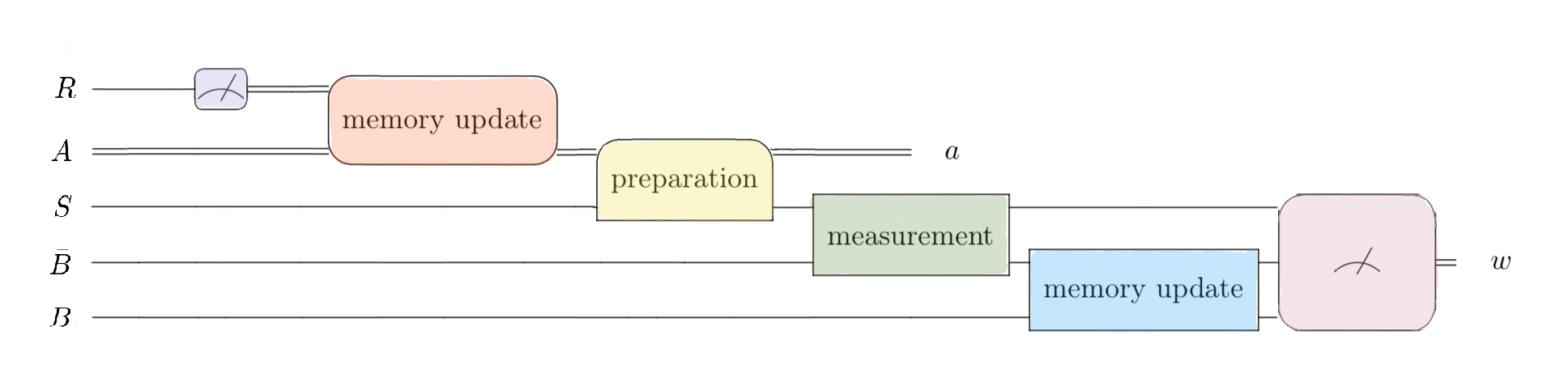}
        \caption{Alice's view: for Alice the measurement outcome $a$ is classical, in the sense that it has a definite value. The same holds of course for the corresponding statements she writes into her memory. However, to make a prediction about the outcome $w$ of the measurement that Wigner applies to Bob's lab, she must treat all parts of the latter as quantum systems.}
        \label{fig:FRA}
    \end{subfigure}
    \\
    \begin{subfigure}{0.9\textwidth}
    \centering
     \includegraphics[scale=.45]{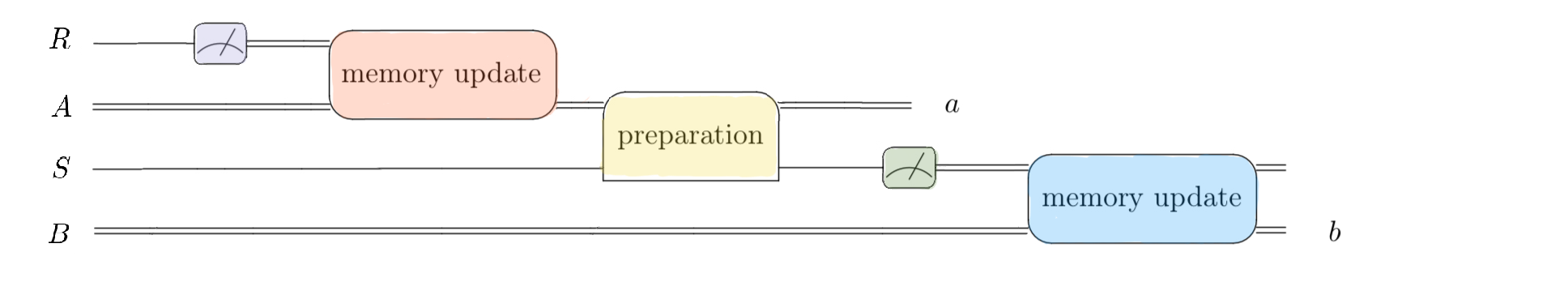}
        \caption{Bob's view: Since Bob is asked to use quantum theory to infer Alice's measurement outcome $a$ depending on his own observation $b$, he should regard both of them as classical.}
        \label{fig:FRB}
    \end{subfigure}
    \\
    \begin{subfigure}{0.9\textwidth}
    \centering
    \includegraphics[scale=.45]{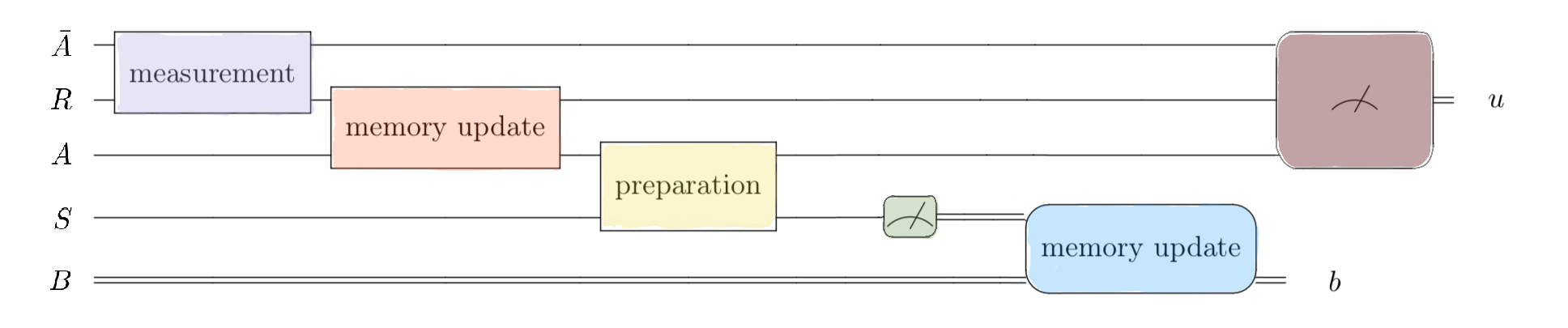}
        \caption{Ursula's view: Ursula applies a measurement on Alice's lab, and she thus has to model its components as quantum systems. Since she is supposed to make a statement about Bob's outcome~$b$, she regards this as a classical value.}
        \label{fig:FRU}
    \end{subfigure}
    \\
     \begin{subfigure}{0.9\textwidth}
     \centering
    \includegraphics[scale=.45]{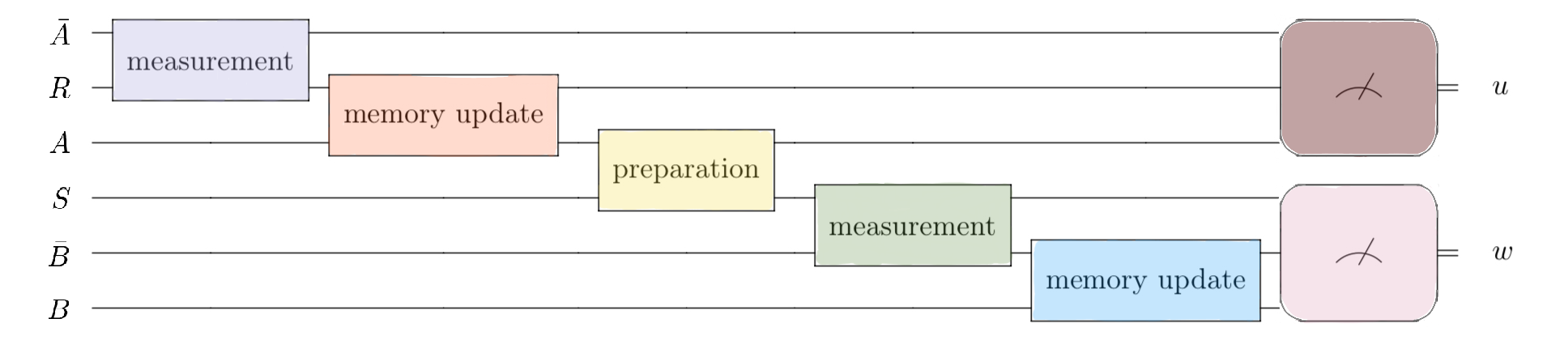}
        \caption{Wigner's view: Wigner does not need to make any statements about the outcomes of Alice and Bob's measurements. He may thus regard their labs as quantum systems, on which his and Ursula's measurements are applied, which yield outcomes $w$ and $u$, respectively.}
        \label{fig:FRW}
    \end{subfigure}
    
\caption{\textbf{Circuit representation of the agents' views on the FR experiment.} The diagrammatic language follows the same scheme as in Figures~\ref{fig:wignercirc} and~\ref{fig:deutschcirc}. Each agent's view focuses on those elements of the experiment that are relevant for their actions and the quantum-theoretic predictions they are supposed to derive. For example, Alice observes $a$ and then uses quantum theory to predict $w$. The later measurement by Ursula (brown box) has according to quantum theory no influence on $w$ and can thus be omitted from Alice's description.} 
\label{fig:FRcirc}
\end{figure}


\subsection{Implications}
\label{subsec:fr-assumptions}

The FR experiment shows that agents, who reason according to the rules $\mathbf{Q}$, $\mathbf{C}$, and $\mathbf{S}$, sometimes run into a contradiction. These rules are thus mutually inconsistent. To better understand this conclusion, it is useful to state more general assumptions from which these reasoning rules are derived. We will also briefly discuss  what the different interpretations have to say about them.

\bigskip

\noindent \textit{\textbf{Assumption $\mathbf{Q}$.} Any agent may describe any system around them using the formalism of quantum theory and, based on this description, infer predictions (or retrodictions) of measurement outcomes using the quantum-mechanical Born rule.}

\smallskip

To turn this assumption into a well-defined reasoning rule, one requires a  quantum-theoretical description of the experiment, e.g., in terms of circuit diagrams. Figure~\ref{fig:FRcirc} shows this for the FR experiment. Just like the diagrams we have seen earlier, each  circuit specifies the viewpoint of one particular agent. It includes those parts that are  relevant for the agent to derive their predictions according to quantum theory. For example, Bob uses his measurement outcome $b$ as a starting point for his inference of the value $a$ observed by Alice. In Bob's view, $a$ and $b$ must thus appear as (classical) measurement outcomes. 

Suppose that an agent's circuit diagram shows that a system in state $\ket{\psi}$ is measured at time $t$ with respect to a basis $\{\ket{e_1}, \ldots, \ket{e_d}\}$. The quantum mechanical Born rule now tells us that the agent will observe outcome $z \in \{1, \ldots, d\}$  with probability $|\inprod{e_{z}}{\psi}|^2$. In fact, for the analysis of the FR experiment it is sufficient to restrict to a special case of this rule. Note that all the predictions by agents that we discussed in Section~\ref{subsec:fr-description} only involve probabilities $0$ or $1$, but nothing in between. In this case the reasoning rule derived from Assumption~$\mathbf{Q}$ admits the following simple form: If  $|\inprod{e_{z}}{\psi}| = 1$ for some~$z = \zeta$, then  the agent can issue the statement
\begin{align} \label{eq_StatementCertainOutcome}
  \text{``I am certain that the outcome $z$ of the measurement at time $t$ is $z=\zeta$.''}
\end{align}

The crux of Assumption~$\mathbf{Q}$ is that no restriction is imposed on the size of systems that it is applicable to. In particular, an agent may describe another agent as a quantum system. For this reason, the assumption is incompatible with certain interpretations of quantum theory. According to the conventional Copenhagen interpretation, for instance, all users of the theory must be above the Heisenberg cut and hence cannot be modelled as quantum systems.   Assumption~$\mathbf{Q}$ is even more explicitly violated by objective collapse theories, which assert that the usual laws of quantum theory are merely  approximations which fail for macroscopic systems. Similar conclusions hold for the ETH approach and consistent histories, which impose a restriction on the set of possible observables, so that certain systems effectively behave classically.

Assumption~$\mathbf{Q}$ also does not hold within Bohmian mechanics. The reason is however different in this case. According to Bohmian mechanics, whether or not a particular measurement outcome occurs is determined by the configuration of the hidden variables rather than the Born rule. The Born rule is hence not a fundamental postulate of Bohmian mechanics, but must be regarded as a derived notion. Its derivation however requires additional assumptions, such as the equilibrium hypothesis (see Section~\ref{subsec:bohmian}). While these assumptions are satisfied in typical situations, they may not be in Wigner's friend-type experiments, where observers are themselves subject to measurements.\footnote{An analysis of the FR experiment within Bohmian mechanics can be found in~\cite{Sudbery2017,LazaroviciHubert2019,Drezet2018}. (Note that~\cite{Drezet2018} also provides a discussion within the Copenhagen interpretation, which is however based on an additional postulate as described in the part on  ``Hadamarded'' agents in Section~\ref{sec:criticism} below. Roughly, it asserts that the conclusions that Alice draws from applying rule~$\mathbf{Q}$ may be rendered invalid by erasing her memory.)\label{ftn:BohmianAnanlysis}}

Conversely, Assumption~$\mathbf{Q}$ is compatible with all other interpretations we have discussed earlier, in particular neo-Copenhagen, many worlds interpretations, RQM, and QBism. According to these interpretations, quantum theory is applicable universally. 

\bigskip

\noindent \textit{\textbf{Assumption $\mathbf{C}$.} An agent can use conclusions  obtained by admitting the view of another agent.}

\smallskip

This assumption may be immediately turned into a reasoning rule for agents. Suppose that two agents, $A$ and $B$, start with the same knowledge about an experiment and reason according to the same set of rules. Assume now that agent~$A$ has established the statement
\begin{align} \label{eq_CInd}
  \text{``I am certain that agent~$B$ is certain that $z=\zeta$ at time $t$.''}
\end{align}
Assumption~$\mathbf{C}$ now allows agent $A$ to conclude
\begin{align}  \label{eq_CDir}
  \text{``I am certain that $z=\zeta$ at time $t$.''}
\end{align}

The implication $\text{\eqref{eq_CInd}} \implies \text{\eqref{eq_CDir}}$ is certainly well motivated. Indeed, if agent $A$ accepted~\eqref{eq_CInd} but rejected~\eqref{eq_CDir}, this would mean that $A$ does not trust in the reasoning rules $B$ is using, which are however the same as $A$'s. Nonetheless, whether or not Assumption~$\mathbf{C}$ is endorsed by the different interpretations is difficult to answer in general, as most of them do not provide receipts for how to deal with scenarios involving more than one single agent.

A notable exception is QBism, which rejects Assumption~$\mathbf{C}$. Rather, QBism holds that an ``agent will have to decide on a case by case basis'' whether or not they adopt another agent's conclusions~\cite{deBrota2020}.   Another exception is the consistent histories interpretation, which relies on the idea that a user of quantum theory must choose a \emph{framework} (see Section~\ref{subsec:consistent}). This framework then restricts the  set of statements that the user is allowed to make. If two users choose different frameworks then their conclusions may be incompatible. In other words, the consistent histories approach explicitly accepts that two users of the theory reach contradictory conclusions, which amounts to denying Assumption~$\mathbf{C}$. 

Assumption~$\mathbf{C}$ is related to the notion of \emph{observer-independent facts}~\cite{Brukner2018}. A measurement outcome is considered observer-independent if different observers can jointly assign a truth value to it.\footnote{Brukner proved the following theorem~\cite{Brukner2018}, which may be regarded as a strengthening of Bell's theorem~\cite{Bell1966}. One of the following assumptions must be wrong: (i)~quantum theory is universally valid (similar to Assumption~$\mathbf{Q}$),  (ii)~agents can freely choose what to measure, (iii)~influences are only local, and (iv)~measurement outcomes are  observer-independent. (See also the further extension of this theorem in~\cite{Cavalcanti}.)\label{ftn:ObserverIndependentFacts}} The assumption of  observer-independent facts is however strictly stronger than Assumption~$\mathbf{C}$. Clearly, if measurement outcomes are observer-independent then $\mathbf{C}$ must hold. However, conversely, $\mathbf{C}$ does not imply that measurement outcomes are observer-independent. A good example illustrating this is the original Wigner's friend experiment. Since Wigner is uncertain about Alice's outcome, $\mathbf{C}$ does not imply anything. But under the assumption of observer-independent facts, Wigner would also need to consider Alice's outcome as a fact, although he is uncertain about its value. 

\bigskip

\noindent \textit{\textbf{Assumption $\mathbf{S}$.} A measurement carried out by an agent has a definitive outcome from the viewpoint of the measuring agent.}

\smallskip

The meaning of this assumption is again best captured in terms of the reasoning rules that it implies. Firstly, $\mathbf{S}$ asserts that if an agent measures a system at a given time $t$ and observes outcome $z=\zeta$ then the agent is entitled to put down the statement ``I am certain that $z=\zeta$ at time $t$.''\footnote{The necessity of this part of the assumption was noted by Bub~\cite{Bub2020}. In fact, taking seriously the idea that all statements must be relative to agents, it may be phrased as follows: If an agent $A$ has established the statement  ``I am certain that agent $B$ observes $z=\zeta$ at time $t$.'' then $A$ can infer the statement ``I am certain that agent $B$ is certain that $z=\zeta$ at time $t$.''\label{ftn:SRemark}} Secondly, the outcome must be definite, in the sense that for any $\zeta' \neq \zeta$ a statement of the form ``I am certain that $z=\zeta$ and I am certain that $z=\zeta'$ at time $t$.'' is disallowed.

Assumption~$\mathbf{S}$ is rather weak. It merely demands that measurement outcomes are defined relative to the measuring agent. It is therefore compatible with all major interpretations of quantum theory, including subjective ones such as QBism or RQM, according to which measurement outcomes do not represent an objective reality, but rather are personal to an agent. For example, applied to the original Wigner's friend thought experiment, the assumption merely asserts that Alice's measurement has a single outcome for her, $a = 0$ or $a=1$. However, for Wigner, this outcome may not exist. 

One may still be worried that Assumption~$\mathbf{S}$ is in conflict with the many-worlds view, according to which a measurement leads to a branching, with a branch associated to each possible outcome. According to this interpretation, Alice may in one branch claim that she observed $a=0$ and in another that she observed $a = 1$. However, crucially, in no branch she would issue a statement of the form ``I observed $a=0$ and $a=1$.'' The reasoning rules derived from~$\mathbf{S}$ are thus compatible with many-worlds interpretations, too.


\section{Attempts to resolve the contradiction}
\label{sec:criticism}
Assumptions $\mathbf{Q}$, $\mathbf{C}$, and $\mathbf{S}$ are contradictory, as we have seen. This is of course not an issue for interpretations that deny the universal validity of quantum theory, notably collapse interpretations. According to them, $\mathbf{Q}$ simply does not hold. Most modern interpretations, however, such as QBism, neo-Copenhagen, and RQM, but also the Everettian many-worlds interpretation and its derivatives, endorse $\mathbf{Q}$. Since these interpretations also subscribe to~$\mathbf{S}$, they must deny $\mathbf{C}$. It thus seems they have to bite the bullet and accept that different agents' statements about the same experiment will sometimes be contradictory.

But maybe there is another way out. Couldn't it be that the contradiction which arises in the FR thought experiment has a different origin? For example, there could be further assumptions, which remained hidden because they appear so natural to us that we wouldn't recognise them as assumptions.\footnote{For example, as rightly pointed out by Anthony Sudbery, the argument in~\cite{Frauchiger2018} uses the assumption that ``If an agent $A$ prepares a system $S$ in a state $\ket{\psi}$, then immediately after the preparation $A$ is certain that $S$ is in the state $\ket{\psi}$''~\cite{Sudbery2019}. This assumption is related to Assumption~$\mathbf{S}$ and is addressed by the formulation given here (see in particular Ftn.~\ref{ftn:SRemark}) together with the understanding that all statements about quantum states are relative to the agent who makes the statement.} In the following we discuss a few possibilities of what these could be. Many of them correspond to points that were highlighted in debates about the FR argument.

\paragraph{Agents are modelled as abstract information-processing systems.} In the FR thought experiment, agents play a crucial role as \emph{users of quantum theory} whose task is to derive predictions. In our analysis so far we basically treated them as information-processing systems, or computers, programmed with the inference rules implied by $\mathbf{Q}$, $\mathbf{C}$, and $\mathbf{S}$, as well as a description of the experiment. They take as input observations and output statements about other (future or past) observations. 

While this rather abstract characterisation of agents was sufficient to describe and analyse the FR experiment, one may argue that it misses other aspects of \emph{agency} that could be physically relevant, too. For example, human agents have consciousness (as pointed out by Wigner~\cite{Wigner1961}), and this may impose constraints on the experiment that are not captured by their modelling as programmable computers.

A resolution of the FR paradox along these lines could thus look as follows. Assumption~$\mathbf{Q}$ is modified in two ways. Firstly,  the notion of agents is tightened so that only humans count as legitimate users of quantum theory. Secondly, the range of validity of quantum theory is restricted to non-human systems. With these modifications, $\mathbf{Q}$ no longer allows a (human) agent to describe another (human) agent as a physical system. This, however, just amounts to giving up Assumption~$\mathbf{Q}$ in its current universal form. 

\paragraph{Agents share common reference frames.} Information is always encoded in physical carriers. Hence, if for instance Alice sends a qubit $S$ to Bob, this qubit must be understood as a physical system, e.g., a particle with a spin degree of freedom that encodes information. But for this encoded information to make sense to both Alice and Bob, they need a common reference frame, which defines for instance the spatial directions up, down, left, or right. Assuming that quantum theory is universal, this reference frame must itself be a quantum system that is accessible to both Alice and Bob. Since Alice and Bob are in isolated labs, however, it is not obvious how to ensure they both can access this system. One may thus try to resolve the contradiction by arguing that it is impossible for Alice and Bob to share a common reference frame. 

Recently, research on quantum reference frames has been gaining momentum~\cite{Angelo2011,Giacomini2019,Castro2019,de2020}, and it has been found that the preparation and measurement of quantum states requires a reference frame of a certain minimum size~\cite{Skotiniotis2017,Lopez2019}.  In the case of a reference system for time, it has been shown however that the conclusions of the FR argument can be maintained even if one models the reference explicitly~\cite{RossiSoares-Pinto2020}. Nonetheless, a full analysis of the FR argument including all necessary reference frames is still pending, and it may well be that the requirement of having such references imposes fundamental constraints on the type and size of quantum systems that can be described by an agent. Ultimately, this would however again mean that one has to reject Assumption~$\mathbf{Q}$.

\paragraph{Agents use rules of logic for their reasoning.} In our analysis of the FR experiment we assumed that the agents reason according to the rules $\mathbf{Q}$, $\mathbf{C}$, and $\mathbf{S}$. But we also assumed, implicitly, that the agents may combine these rules by logical reasoning. For example, Bob establishes that if Alice had observed $a=0$ then he would not have observed $b=1$. From this he infers that, if he observed $b=1$ then Alice could not have observed $a=0$. In other words, Bob applies the standard logical reasoning step that if $U \implies (\mathrm{not} \, V)$ then $V \implies (\mathrm{not} \, U)$.

Instead of rejecting one of the rules $\mathbf{Q}$, $\mathbf{C}$, and $\mathbf{S}$, one may thus as well reject the idea that the agents are entitled to apply standard logical reasoning. Disallowing the use of standard logic, however, appears to be a more severe step than giving up, for instance, rule~$\mathbf{C}$. It is even unclear whether rules like $\mathbf{C}$ or $\mathbf{S}$ can keep their intended meaning if logic is abandoned. 

\paragraph{The agents' statements are probabilistic.}  In the FR thought experiment, all statements by agents are of the form ``I am certain that ...''. One may thus be worried that the conclusions of the argument strongly depend on the interpretation of the term ``being certain'', which we haven't defined however. In QBism, for instance, one could take it to mean that the agent assigns probability~$1$ to the corresponding statement. Such a probabilistic interpretation may then help resolving the paradox.

However, as pointed out by Bub~\cite{Bub2020}, the contradiction arising in the FR thought experiment is largely independent of how one interprets the term ``certainty'', provided that the interpretation is compatible with the reasoning rules $\mathbf{Q}$, $\mathbf{C}$, and $\mathbf{S}$. In other words, $\mathbf{Q}$, $\mathbf{C}$, and $\mathbf{S}$ can be themselves understood as the rules defining ``certainty''. Hence, if one wishes to resolve the paradox by a particular interpretation of  the term then one has to choose one that violates at least one of these rules. The contradiction between them thus remains. 


\paragraph{Agents are ``Hadamarded''.} In the Wigner-Deutsch experiment, Alice's lab is subject to a final measurement with respect to a superposition state $\ket{\text{lab}_+}_{R A \bar{A}}$ as defined by~\eqref{eq_AliceSuperposition}, and similar measurements also occur in the FR experiment. As indicated earlier, such measurements may result in a complete erasure of the agents' memory, or, as Scott Aaronson put it, they are akin to a ``Hadamarding'' of the agents' brains~\cite{Aaronson2018}. Figure~\ref{fig:Hadamard} explains why this is the case and where the terminology comes from.

In a much-noticed comment~\cite{Aaronson2018} about the FR paradox, Aaronson argued that, due the detrimental effect of the measurements that are applied to Alice and Bob's labs, these agents can no longer remember the statements they derived. His conclusion was that the contradiction arising in the FR thought experiment may be avoided simply by declaring that an agent's  inferences obtained from~$\mathbf{Q}$ become void once the agent has lost their memory. (A similar argument has been given in~\cite{Drezet2018}, see Ftn.~\ref{ftn:BohmianAnanlysis}, and in~\cite{Healey2018}.)

\begin{figure}
\centering
    \begin{subfigure}[t]{0.45\textwidth}
     \begin{align*}
       \Qcircuit @C=.7em @R=1em {
         \lstick{L} & \qw & \gate{H}  & \ctrl{1} & \gate{H} & \qw  \\
         \lstick{\ket{0}} & \qw & \qw  & \targ & \qw & \qw & \rstick{w} \\
        }
    \end{align*}
        \caption{The measurement of $L$ with respect to the dual basis $\{\ket{+}, \ket{-}\}$ may be implemented by first applying a Hadamard gate, $H$, which by definition maps $\ket{+}$ to $\ket{0}$ and $\ket{-}$ to $\ket{1}$, then measuring with respect to the computational basis $\{\ket{0}, \ket{1}\}$, and then transforming back with $H$ (which is its own inverse). The computational basis measurement is furthermore modelled as a \emph{controlled not} operation, which effectively copies the value in $L$ into the wire carrying the result.}
        \label{fig:Hadamard-a}
    \end{subfigure}
    \hfill
    \begin{subfigure}[t]{0.45\textwidth}
    \begin{align*}
       \Qcircuit @C=.7em @R=1em {
         \lstick{L} & \qw & \qw & \targ & \qw & \qw  \\
         \lstick{\ket{0}} & \qw & \gate{H}  & \ctrl{-1} & \gate{H} & \qw & \rstick{w} \\
        }
    \end{align*}
        \caption{This circuit can be shown to be equivalent to that of~(a) --- the unitary evolution of the two wires is precisely the same. This representation however emphasises the detrimental impact that the measurement has on~$L$. The state of the additional wire after the first Hadamard gate is $\ket{+}$, i.e., an equal superposition of $\ket{0}$ and $\ket{1}$. The controlled not operation thus adds a random bit to~$L$, i.e., the information contained in the lab gets completely randomised.}
        \label{fig:Hadamard-b}
    \end{subfigure}
\caption{\textbf{Hadamarding measurements.} Suppose that the state of Alice's lab is represented by a qubit, $L$, with the \emph{computational basis states} $\ket{0}$ and $\ket{1}$ corresponding to $\ket{\text{lab}_0}_{R A \bar{A}}$ and $\ket{\text{lab}_1}_{R A \bar{A}}$, respectively as defined by~\eqref{eq_AliceMeasurement}. Measuring the lab with respect to superposition states like~\eqref{eq_AliceSuperposition} then corresponds to measuring $L$ with respect to the \emph{dual basis states} $\ket{\pm} = \smash{\sqrt{\frac{1}{2}}(\ket{0} \pm \ket{1})}$. This measurement may be represented by a circuit diagram with one wire for $L$ and one additional wire where the measurement outcome $w$ is stored. The circuit consists of particular operations, known as \emph{Hadamard gates}~\cite{Nielsen2009}, hence the name ``Hadamarding''.}
\label{fig:Hadamard}
\end{figure}

Such a declaration would however be hard to justify. The timing of the individual steps in the FR experiment is chosen such that no agent is required to remember statements (let alone to derive them or take any other actions) after a measurement has been applied to their lab.  Hence, if one  wishes to use Aaronson's argument to resolve the FR paradox, one would need to postulate, for instance,  that the statement about the outcome $w$ of Wigner's measurement that Alice derives at time $t=1$ using rule~$\bf{Q}$, and which is known to Bob at time $t=2$, is rendered invalid by the measurement that Ursula applies to Alice later at time $t=3$. This amounts to dismissing quantum-theoretic predictions that an agent makes at present on the grounds that the agent's brain will at some later time  deteriorate.

\bigskip

We have thus seen various proposals for how one could resolve the FR paradox. Some of them amount to giving up some of the most basic assumptions that underlie physical reasoning (like disallowing the use of standard logic, or postulating that future actions on an agent's brain can invalidate conclusions drawn by the agent in the past). The others basically correspond to rejecting one of the explicit assumptions, $\mathbf{Q}$, $\mathbf{C}$, or $\mathbf{S}$.

\section{Conclusion and outlook}
\label{sec:discussion}
Wigner's thought experiment points to a conflict between the idea that quantum theory is a universally valid theory and our experience that measurements yield definite outcomes. The nature of this conflict has been sharpened by the later extended variants of the thought experiment. The conclusion may be phrased as a theorem, which by itself is  interpretation-independent: three reasonably sounding assumptions, namely~$\mathbf{Q}$ (quantum theory is universally valid), $\mathbf{C}$ (the views of different agents is mutually consistent), and $\mathbf{S}$ (a measurement has a single outcome for the measuring agent), are contradictory. But, clearly, physics should not be contradictory, hence at least one of the three assumptions must be inaccurate.

The three assumptions are linked to specific reasoning rules that users of quantum theory may apply when deriving predictions about the outcomes of an experiment. While these reasoning rules are undisputed in the context of the usual experiments we are carrying out in our labs, their use is no longer straightforward in special settings like the FR experiment. Here the predictions depend on our choice of which assumptions we keep or modify. Making this choice is thus a physics problem.

This is the point where interpretations come into play. None of the major interpretations of quantum theory dismisses Assumption~$\mathbf{S}$. Taking thus $\mathbf{S}$ for granted, one is basically left with a decision between $\mathbf{Q}$ and $\mathbf{C}$, and the interpretations separate into two categories. One consists of interpretations such as  conventional Copenhagen or collapse theories that reject~$\mathbf{Q}$. According to them, quantum theory is not universally valid --- rather certain parts of the world belong to a classical domain and are thus not accurately described by the theory. The other category includes interpretations like neo-Copenhagen, QBism, or relational quantum mechanics, according to which  quantum theory is applicable to any physical system, so that~$\mathbf{Q}$ holds. However, these interpretations then have to dismiss Assumption~$\mathbf{C}$, i.e., statements made by different agents will not always be mutually consistent. 

Neither of the two options seems to be satisfactory. Any interpretation that rejects~$\mathbf{Q}$ should include a clear-cut criterion that defines the type of systems to which the quantum formalism still applies. However, except for certain collapse theories, none of the interpretations in this category provide such a criterion.\footnote{In the case of the conventional Copenhagen interpretation, for instance, the criterion would need to state where to put the Heisenberg cut that separates the quantum from the classical domain. In the case of the ETH approach, one would need a receipt for how to choose the basic set of observables.} Conversely, giving up the consistency assumption~$\mathbf{C}$ without replacement amounts to depriving agents from their ability to reason about each other --- a rather solipsistic view. Any interpretation that dismisses~$\mathbf{C}$ should  thus substitute it with other means for agents to compare their individual views. While this problem has been largely ignored until recently (but see~\cite{deBrota2020,Narasimhachar2020}), one can be optimistic that future research, inspired by (real or thought) experiments, will soon shed more light on it. 

\paragraph{Real experiments.}

Experiments over more than a century have confirmed the predictions of quantum theory  to higher and higher accuracy and for larger and larger systems. Many physicists thus take it for granted that the domain of applicability of quantum theory ranges to macroscopic systems.\footnote{This conviction goes that far that physicists regularly apply quantum theory to describe objects of astronomical or cosmological scale. Examples include discussions around the black hole information paradox, as well as arguments on the origin of the microwave background radiation in cosmology.} Considering the thought experiments discussed here, it may however be worth redirecting our search. Recall that we used  Assumption~$\bf{Q}$ to justify that agents, i.e., the users of quantum theory, are themselves subject to the laws of the theory. Hence, rather than testing larger and larger systems, it could make sense to test whether systems that can reasonably count as agents are still accurately described by quantum theory.

Quite obviously, such a test cannot be carried out with human agents. But  for the purpose of our arguments, an agent can be any system that is able to use quantum theory. The agents may thus be substituted by computers.  The requirement of the Wigner-Deutsch experiment or the FR experiment that agents are in isolated labs would then correspond to demanding that the information-carrying degrees of freedom of the computers are shielded from their environment. Note that this is necessarily the case for quantum computers, i.e., they would be natural devices to simulate agents in Wigner's friend type experiments.\footnote{A main challenge in building a quantum computer is precisely to protect the information stored in them from decoherence, which means that the information-carrying degrees of freedom must be well isolated.} Taking this further, quantum computers may become a valuable experimental tool in research on quantum foundations.

\paragraph{G\"odel's theorem and further thought experiments.} The FR argument appears to have some similarities to G{\"o}del's incompleteness result~\cite{Godel1931}. Indeed, both of them use a self-referential statement (in the case of the FR argument, an indirectly self-referential chain of statements) to arrive at their respective conclusions. However, while incompleteness theorems show that it is \emph{impossible to prove} that a given theory is consistent within the theory itself, the FR argument proves that the rule system consisting of $\textbf{Q}$, $\textbf{C}$, and $\textbf{S}$ \textit{is not} consistent. In this sense, the FR argument yields a converse statement to G\"odel's theorem.

One may also ask whether the conclusions obtained from the thought experiments presented here are specific to quantum theory, or whether they hold more generally. In \cite{Vilasini2019}, it was shown that, within the framework of \emph{generalised probability theories} \cite{Hardy01, Barrett07, Gross2010}, which generalise the operational aspect of quantum theory, this is indeed the case for an experiment analogous to the FR thought experiment. Conversely, one may also apply classical multi-agent modal logic~\cite{Fagin2003} to investigate Wigner's friend type experiments.  This however turned out to be challenging, for these classical frameworks rely on assumptions that appear unnatural when reasoning within quantum theory~\cite{NL2018, Boge2019, Fraser2020}. 

\bigskip

We conclude by noting that, although quantum theory is phenomenologically extremely accurate, it appears to be difficult to reconcile with other naturally sounding assumptions. Specifically, if we demand that the views of different agents should be mutually consistent, we run into contradictions. The situation vaguely resembles that of classical mechanics before the invention of relativity theory: at that time, classical mechanics was a phenomenologically accurate theory, too. However, thought experiments involving observers that move at (or close to) the speed of light exhibited contradictions, which eventually forced us to abandon it as a fundamental theory of Nature. Hence, if history can be of any guidance here, it may be worth putting more effort into the study of thought experiments to explore the limits of  quantum theory and identify possible inconsistencies. It could well be that Wigner's friend still has a story to tell.

\begin{acknowledgements}
NN and RR acknowledge support from the Swiss
National Science Foundation through 
 project No.\ $200020\_165843$, through 
the National Centre of
Competence in Research \emph{Quantum Science and Technology}
(QSIT) and the National Centre of Competence in Research \textit{SwissMAP --- The Mathematics of Physics}.
\end{acknowledgements}

\bibliographystyle{unsrtnat}

\bibliography{lit}

\end{document}